\documentclass[prb,twocolumn,showpacs]{revtex4}
\usepackage{graphicx}
\usepackage{dcolumn}
\usepackage{bm}
\usepackage{amsmath}

\begin{document}

\title{Variational description of the ground state of the repulsive two-dimensional Hubbard model in terms of
 nonorthogonal symmetry-projected Slater determinants
}

\author{R. Rodr\'{\i}guez-Guzm\'an$^{1,2}$, Carlos A. Jim\'enez-Hoyos$^{1}$ and Gustavo E. Scuseria$^{1,2}$}

\affiliation{
$^{1}$ Department of Chemistry, Rice University, Houston, Texas 77005, USA
\\
$^{2}$ Department of  Physics and Astronomy, Rice University, Houston, Texas 77005, USA
}

\date{\today}

\begin{abstract}
The few determinant (FED) methodology, introduced in our previous works [Phys. Rev. B {\bf{87}}, 235129 (2013)
and Phys. Rev. B {\bf{89}}, 195109 (2014)] for one-dimensional (1D) lattices, is here adapted for the repulsive two-dimensional
Hubbard model at half-filling and with finite doping fractions. Within this configuration
mixing scheme, a given ground state with well defined
spin and space group quantum numbers, is
expanded in terms of a nonorthogonal symmetry-projected basis determined through chains of
variation-after-projection calculations. The results obtained for the ground state and correlation energies
of half-filled and doped 4 $\times$ 4, 6 $\times$ 6, 8 $\times$ 8,
and 10 $\times$ 10 lattices, as well as momentum distributions and spin-spin correlation functions
in small lattices, compare well with those obtained using other state-of-the-art approximations.
The structure of the intrinsic determinants resulting
from the variational strategy is interpreted in terms of defects that encode information on the
basic units of quantum fluctuations in the considered 2D systems. The varying nature
of the underlying quantum fluctuations, reflected in a transition to a stripe regime for increasing
onsite repulsions, is discussed using
the intrinsic determinants belonging to a 16 $\times$ 4 lattice with 56 electrons.
Such a transition is further illustrated by computing spin-spin
and charge-charge correlation functions with the corresponding multireference FED wave functions.
In good agreement with previous studies, the analysis of the pairing
correlation functions reveals a weak enhancement
of the extended $s$-wave and d$_{x^{2}-y^{2}}$ pairing modes. Given the quality of results here reported
together with those previously obtained for 1D lattices and the parallelization properties of the FED scheme,
we believe that symmetry projection techniques
are very well suited for building ground state wave functions
of correlated electronic systems, regardless of their dimensionality.

\end{abstract}


\maketitle

\section{Introduction.}
\label{INTRO}

Due to its challenging complexity, the description
of low-dimensional correlated electronic systems still
represents an open problem in condensed matter physics. \cite{Dagotto-review,Dagotto-Rev-Mod-Physics-2013}
In particular, their quantum correlation effects
can exhibit unconventional features. A typical example, is
the spin-charge separation \cite{text-Hubbard-1D,Voit,Ogata-Shiba} in the
strong interaction regime of the
one-dimensional (1D) Hubbard model. \cite{Hubbard-model_def1,Gebhard-1}
Angle-resolved photoemision
spectroscopy studies also reveal a complex pattern
of spin-charge coupling/decoupling in both the
1D and two-dimensional (2D)
cases in the weak and intermediate-to-strong
interaction regimes.  \cite{Kim,KMShen}
How to account for these, and many more, quantum correlation
effects in the simplest possible way has become a driving
force for developing  theoretical approximations that could
complement already existing
state-of-the-art methods like
exact diagonalization \cite{Dagotto-review,Lanczos-Fano} (ED),  quantum
(QMC)
and variational (VMC)
Monte Carlo, \cite{Raedt-MC,Nightingale,Shiwei-QMC-Symmetry,Neuscamman-2012} coupled
cluster, \cite{Bishop-1,Bishop-2} variational
reduced-density-matrix, \cite{Hammond} density matrix renormalization
group, \cite{DMRG-White,Scholl-RMP} matrix product and tensor network states,
\cite{Scholl-AP,GChan,TNPS-1,TNPS-2,Vidal}
as well as quantum embedding approaches.
\cite{Zgid,maier2005,stanescu2006,DMFT-1,moukouri2001,huscroft2001,aryanpour2003,DVP-2,Knizia-Chan,Irek-DMET-paper1,Irek-DMET-paper2,DMET-honey}
 All these methods have already been applied to
 Hubbard-like 1D and/or 2D models with variable
degree of success.

The exact Bethe-ansatz solution to the 1D Hubbard model is well known. \cite{BETHE,LIEB}
Because of this, the model has been frequently used as a testing ground for
several theoretical frameworks. However, an intuitive
physical picture of the basic
units of quantum fluctuations in the considered 1D
systems  has  remained
an open issue within several  approximations. In recent years, both single reference
(SR) and multireference (MR) symmetry-projected
approximations, \cite{Tomita-1,Tomita-2,Tomita-3,Tomita-2011PRB,Carlos-Hubbard-1D,Rayner-2D-Hubbard-PRB-2012,non-unitary-paper-Carlos,
rayner-Hubbard-1D-FED2013,rayner-Hubbard-1D-FED2014,Carlos-Rayner-Gustavo-VAMPIR-molecules,Carlos-Rayner-Gustavo-FED-molecules,Laimis-paper,
PQT-reference-1,PQT-reference-2,PQT-reference-3,Juillet}
routinely used in nuclear structure physics, \cite{rs,Carlo-review,Schmid-Gruemmer-1984,Rayner-Carlo-CM-1,Rayner-Carlo-CM-2,rayner-GCM-paper,rayner-GCM-parity}
have been applied to
describe correlated electronic systems. It has been shown
that MR schemes like the
Resonating Hartree-Fock
\cite{Tomita-1,Tomita-2,Tomita-3,Tomita-2011PRB,Fukutome-original-RSHF,Yamamoto-1,Yamamoto-2,Ikawa-1993}
(ResHF) and the Few Determinant
\cite{Carlo-review,Schmid-Gruemmer-1984,rayner-Hubbard-1D-FED2013,rayner-Hubbard-1D-FED2014,Carlos-Rayner-Gustavo-VAMPIR-molecules,Carlos-Rayner-Gustavo-FED-molecules}
(FED) ones, provide a reasonable
description of the ground state energies of half-filled and doped 1D Hubbard lattices
but also  account for the main physical trends in correlation functions, momentum distributions, spectral
functions, and density of states.
\cite{Tomita-1,rayner-Hubbard-1D-FED2013,rayner-Hubbard-1D-FED2014}
In addition, within these approaches, one is left with
a simple physical picture
in which the basic units of quantum fluctuations  in 1D  lattices can be mainly associated with
structural defects in the (intrinsic) Slater determinants resulting from the
corresponding optimizations. \cite{Tomita-1,rayner-Hubbard-1D-FED2013,rayner-Hubbard-1D-FED2014}

The situation is more involved in the case of the 2D Hubbard model for which no general (exact) solution is known.
Such a model has received considerable attention since the discovery of high-T$_{C}$ superconductors \cite{HTCSC-1}
and has also become the target for theorists applying many-body methodologies.
According to Anderson's proposal, \cite{Anderson-1}
2D Hubbard is considered a potential model for describing the essential physics in the
cuprates. With intensive analytic and numerical studies, \cite{Theory-strong}
some aspects of the phase diagram have been understood.
\cite{Dagotto-review} However, many basic features remain controversial. For example, while
it is accepted that the onsite interaction strength drives a Mott transition
to an insulator at half-filling, \cite{Hirsch-1,Hirsch-2,White-AF-HF,Furukawa}
it is much more difficult to accurately describe
what happens to the antiferromagnetic order when the system is doped.

From the experimental point
of view, cold atoms in optical lattices offer potential direct simulations of  Hubbard-like
models. \cite{optical-1} On the other hand, such
models are valuable tools to study the properties of graphene \cite{CastroNeto-review}
and their (multiorbital) extensions \cite{Su-Fe-Dagotto}  have already
provided  insight into the
interplay between electronic correlations and doping in the parent states of
high-T$_{C}$ iron-based superconductors.  \cite{sup-Fe,Stewart-Review}
Futhermore, both the colossal magnetic resistance and large
thermopower has attracted considerable attention.
\cite{Science-Dagotto,Otha}
In addition, a fascinating effort to understand exotic spin liquid phases
in the ground and low-lying excited states of some 2D systems is bringing  new light into
the complexity of the associated many-electron problems and the theoretical tools used
in their description. \cite{SL-1,SL-2,SL-3}
We believe that the previous examples  illustrate the need
to explore new avenues for describing quantum correlation effects in
low-dimensional electronic systems, especially those approximations that are
potentially not restricted by the dimensionality of the considered lattices.

The 2D Hubbard model has already been considered with symmetry projection tools in
a previous work \cite{Rayner-2D-Hubbard-PRB-2012}
using a variation-after-projection (VAP) approach. \cite{rs} For the
considered  lattices, it has been shown that such an approach accurately describes
both ground and low-lying excited states, with well defined quantum numbers, on an equal footing.
The comparison with other state-of-the-art
approximations revealed that the method does account for the most
relevant correlations including a basic
quantum mechanical fingerprint as the
low-lying spectrum of the 6 $\times$ 6 lattice, which is out of reach of
ED calculations. Symmetry-projected methods also provide a well-controlled
ansatz to compute both spectral functions and density of states. However, despite being more
sophisticated than SR methods, \cite{Carlos-Hubbard-1D,PQT-reference-1,PQT-reference-2,PQT-reference-3,Juillet}
our scheme still essentially relied on the
description of a given ground and/or excited
state in terms of a single symmetry-projected configuration (or component).
This certainly limits the amount of correlations that can be described
in the ground and excited states of nuclear, \cite{Carlo-review}
 condensed mater, \cite{Rayner-2D-Hubbard-PRB-2012} and quantum chemistry systems.
\cite{Carlos-Rayner-Gustavo-VAMPIR-molecules}
In this study, we present results that go beyond such a single configuration
and benchmark a MR method, i.e., the FED approach, \cite{rayner-Hubbard-1D-FED2013}
further including the breaking and restoration of the full space group
symmetry, which was not included in our previous work.

%
%
\begin{figure}
\includegraphics[width=0.50\textwidth]{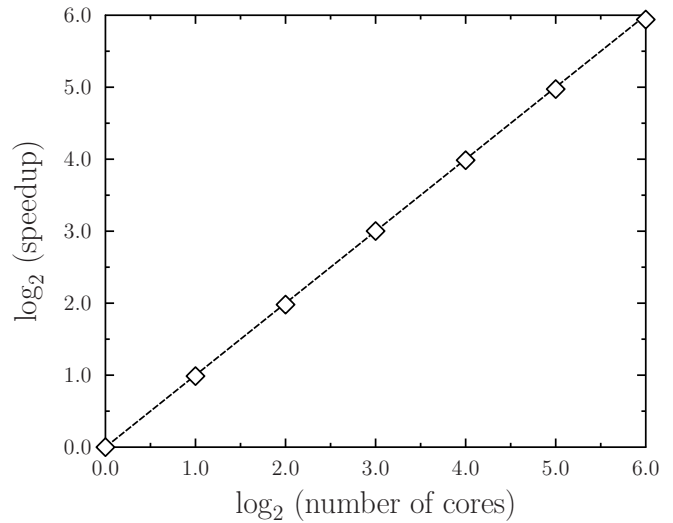}
\caption{Speedup of a typical UHF-FED calculation for a half-filled
12 $\times$ 12 lattice at U=4t. The origin of the plot refers to a calculation with
1024 cores. The largest calculation uses 65,536 processors. All calculations
have been performed at the Titan computational facility, Oak Ridge National Laboratory, Center
for Computational Sciences.
}
\label{Fig1}
\end{figure}
%
%

The key idea of the FED approach
\cite{Carlo-review,rayner-Hubbard-1D-FED2013,rayner-Hubbard-1D-FED2014,Carlos-Rayner-Gustavo-FED-molecules}
is to consider a set of symmetry-broken
Hartree-Fock (HF) states $| {\cal{D}}^{i} \rangle$
which are used to build, via chains of Ritz-variational calculations, \cite{Blaizot-Ripka}
a correlated nonorthogonal basis
of  $n$ symmetry-projected configurations $\hat{P}^{\Theta} | {\cal{D}}^{i} \rangle$,
with  $\hat{P}^{\Theta}$ being a projection operator
(see, Sec. \ref{Theory})
 characterized by the quantum numbers $\Theta$
associated with the irreducible representations of the symmetry groups
under consideration. The FED wave function is simply a
variationally optimized expansion in terms of these $n$ symmetry-projected  states.
Let us stress that FED is a
VAP scheme,  within which the
intrinsic states $| {\cal{D}}^{i} \rangle$ are
always
optimized
in the presence of the projection operator $\hat{P}^{\Theta}$. This is what  brings a different
structure (i.e., defects) in  each of the  determinants $| {\cal{D}}^{i} \rangle$
as compared with the standard HF ones. \cite{StuberPaldus,HFclassification}  These intrinsic
states  $| {\cal{D}}^{i} \rangle$ are
 optimized one-at-a-time
within the FED approach. \cite{rayner-Hubbard-1D-FED2013} A simultaneous
optimization of all the transformations ${\cal{D}}^{i}$ can become quite demanding in situations
where large expansions in terms of nonorthogonal symmetry-projected configurations are required. \cite{rayner-Hubbard-1D-FED2014}
In fact, it is the FED VAP strategy what allows us to reach expansions larger than those
possible within the ResHF scheme,
as well as to alleviate our numerical effort in calculations based on the most general
HF intrinsic states
that require full three-dimensional spin projection. The reason for
this is quite simple: in a ResHF optimization ${\cal{O}}(n^{2})$ Hamiltonian
and norm kernels have to be recomputed at every iteration
while only ${\cal{O}}(n)$ kernels are required
in an efficient implementation of FED.
Note, however, that we keep the acronym FED just to remain consistent
with the literature; there is no need for the FED expansion to be short,
as its name would imply, although it is certainly a desirable feature. Even in the
case of a SR expansion (i.e., $n=1$), the
wave function is, via the projection operator $\hat{P}^{\Theta}$, already
multideterminantal in nature, \cite{Rayner-2D-Hubbard-PRB-2012}  making  it
a high-quality trial state for the
constrained-path QMC (CPQMC) approximation.  \cite{PaperwithShiwei}
Last but not least, small vibrations around symmetry-projected
mean-fields (i.e., symmetry-projected Tamm-Dancoff and random phase
approximations) can be consistently formulated both at the SR
and MR levels. \cite{RPA-Schmid,RPA-Nishiyama} Such an approximation has been
recently used  to access a large number of excited states required to compute
optical conductivity in lattice models.
\cite{opticalCon-1,opticalCon-2,opticalCon-3}
Results will be presented in a forthcoming publication. \cite{RPA-Rayner}

%
%
\begin{figure}
\includegraphics[width=0.45\textwidth]{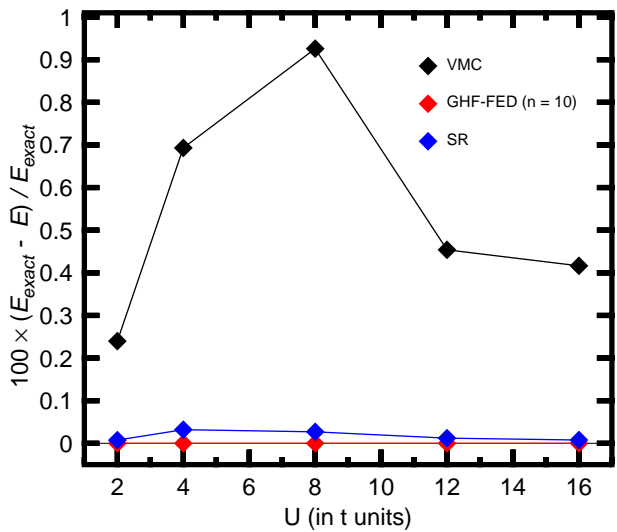}
\caption{Relative energy errors obtained with the GHF-FED approach
based on $n=10$ transformations (red diamonds) are compared with
the ones obtained within SR calculations (blue diamonds)
as well as with VMC results  based on a CPS-Pfaffian ansatz
\cite{Neuscamman-2012}
 (black diamonds). Results
are shown for a half-filled 4 $\times$ 4 lattice and onsite
 interactions of U=2t, 4t, 8t, 12t and 16t. For more details, see the main text.
}
\label{Fig2}
\end{figure}
%
%

In this paper we adapt the FED methodology, introduced in our previous studies of
1D Hubbard lattices, \cite{rayner-Hubbard-1D-FED2013,rayner-Hubbard-1D-FED2014}
to the  half-filled and doped 2D Hubbard model. Our main goal here
is not to be exhaustive but rather to test the method's performance
via benchmark calculations on a selected set of illustrative examples. It will
be shown below that the FED approach provides accurate correlated ground
state wave functions with well defined quantum numbers for 2D systems.
For completeness, we briefly describe the key ingredients of our MR approach
and set our notation in Sec. \ref{Theory}.
For a more detailed account, the reader is referred to our previous work. \cite{rayner-Hubbard-1D-FED2013}
We also illustrate the computational performance of our scheme.
Our calculations are discussed in Sec. \ref{results}.
In Sec. \ref{results-gs-corr-energies}, we compare  ground state and correlation
energies with those obtained using
other state-of-the-art approaches. We demonstrate the feasibility
of FED calculations on half-filled and doped 2D lattices with
16, 36, 64, and 100 sites. Most of the calculations have been carried out
for onsite interactions U=2t, 4t, 8t, and 12t, taken as representatives
of the weak, intermediate-to-strong, and strong interaction regimes,
respectively. We also discuss the
dependence of the predicted correlation energies on the
number of basis states used in the corresponding FED expansions, as well as
the structure of the intrinsic determinants resulting from the
VAP procedure. Having discussed the energetic quality of our
states, we turn our attention in Sec. \ref{Corre-functions} to momentum
distributions and correlation functions. There, we first calibrate the quality of
our results in  a small 4 $\times$ 4 lattice
with N$_{e}$=14 electrons. Subsequently, we consider the momentum
distribution, spin-spin (SSCF), charge-charge (CCCF) and pairing
(PCF) correlation functions in the case of
a 16 $\times$ 4 lattice with N$_{e}$=56 electrons. We show
how for increasing U values our MR ansatz captures the transition
to the stripe regime predicted with other theoretical tools.
Finally, Sec. \ref{CONCLU} is devoted to concluding remarks and work perspectives.

\section{Theoretical framework}
\label{Theory}

We consider the following one-band version of the 2D Hubbard
Hamiltonian \cite{Hubbard-model_def1,Gebhard-1}

\begin{eqnarray} \label{HAM-hubbard1D}
\hat{H} =
-t \sum_{{\bf{j}},\sigma}
\Big \{
\hat{c}_{{\bf{j}}+{\bf{x}} \sigma}^{\dagger} \hat{c}_{ {\bf{j}}\sigma}
+
\hat{c}_{{\bf{j}}+{\bf{y}} \sigma}^{\dagger} \hat{c}_{ {\bf{j}}\sigma}
+ h.c
\Big \}
+
U \sum_{{\bf{j}}} \hat{n}_{{\bf{j}} \uparrow} \hat{n}_{{\bf{j}} \downarrow}
\end{eqnarray}
where the first term represents nearest-neighbor
hopping  (t $>$ 0) with hopping vectors ${\bf{x}}$=(1,0) and
${\bf{y}}$=(0,1), and the second term is  the  onsite  interaction.
In this work, we concentrate on the repulsive sector of the Hubbard
model, i.e., U $>$ 0.
The fermionic operators
$\hat{c}_{{\bf{j}} \sigma}^{\dagger}$ and $\hat{c}_{{\bf{j}} \sigma}$
create and destroy an electron with spin-projection
$\sigma= \pm 1/2$ (also denoted as $\sigma= \uparrow, \downarrow$) along
 an arbitrary chosen
quantization axis on a lattice site
${\bf{j}}$=(j$_{x}$,j$_{y}$).
The  operators
$\hat{n}_{{\bf{j}} \sigma}$ = $\hat{c}_{{\bf{j}} \sigma}^{\dagger} \hat{c}_{{\bf{j}} \sigma}$
 are the local number operators.
 Here, and in what follows, the lattice indices run as
 j$_{x}$=1, $\dots$, N$_{x}$ and j$_{y}$=1, $\dots$, N$_{y}$
with  N$_{x}$ and N$_{y}$ being the number of sites along the x
and y directions, respectively. We assume periodic boundary
 conditions along both directions as well as
a lattice  spacing $\Delta$=1.

\begin{table*}
\label{Table1}
\caption{Ground state energies per site (in t units)  obtained with the GHF-FED  scheme based on n nonorthogonal
symmetry-projected configurations [Eq.(\ref{FED-state-general})] for the 4 $\times$ 4 lattice at different doping fractions
and U=4t, 8t and 12t are compared with those
obtained within the constrained-path (CPQMC) and release-constraint (RCQMC) QMC
approaches, based on trial CASSCF wave functions with symmetries, as well as with those
obtained via exact diagonalization (ED) calculations.
\cite{Shiwei-QMC-Symmetry}
For each configuration
the corresponding set of symmetry quantum numbers $\Theta$
[in all cases ${\bf{k}}$=(0,0)]
is also given in the table.
For more details, see the main text.
}
\begin{tabular}{cccccccccccccccccccccccc}
\hline
\\
 $U/t$    &  &$N_{e}$  & & & & $\Theta$  & & & & & & CPQMC	               & & & & RCQMC		& & &  & GHF-FED[n]	  &  & &  ED \\
\\
\hline
\\
 4       &  &  4	 & & & & $^{1}$B$_{1}$	   & & & & & & -0.72094(1)     & & & & -0.72063(1)	& & &  & -0.72064[n=1]    &  & &  -0.72064 \\	
\\
 8       &  &  4	 & & & & $^{1}$B$_{1}$	   & & & & & & -0.7082(1)      & & & & -0.7075(2)	& & &  & -0.7076[n=1]	  &  & &  -0.7076 \\
\\
 12      &  &  4	 & & & & $^{1}$B$_{1}$	   & & & & & & -0.7010(1)      & & & & -0.7002(3)	& & &  & -0.7003[n=1]	  &  & &  -0.7003   \\
\\
 4       &  &  8	 & & & & $^{1}$B$_{1}$	   & & & & & & -1.09693(2)     & & & & -1.09597(6)	& & &  & -1.09591[n=10]   &  & &  -1.09593   \\
\\
 8       &  &  8	 & & & & $^{1}$B$_{1}$	   & & & & & & -1.0307(1)      & & & & -1.0282(2)	& & &  & -1.0288[n=20]    &  & &  -1.0288    \\
\\
 12      &  &  8	 & & & & $^{1}$B$_{1}$	   & & & & & & -0.9962(1)      & & & & -0.9940(3)	& & &  & -0.9939[n=10]    &  & &  -0.9941    \\
\\
 4       &  &  10	 & & & & $^{1}$A$_{1}$	   & & & & & & -1.22368(2)     & & & & -1.22380(4)	& & &  & -1.22380[n=20]   &  & & -1.22381	  \\
\\
 8       &  &  10	 & & & & $^{1}$A$_{1}$	   & & & & & & -1.0948(1)      & & & & -1.0942(2)	& & &  & -1.0942[n=10]    &  & & -1.0944      \\
\\
 12      &  &  10	 & & & & $^{1}$A$_{1}$	   & & & & & & -1.0292(1)      & & & & -1.0278(4)	& & &  & -1.0283[n=40]    &  & & -1.0284      \\
\\
 4       &  &  12	 & & & & $^{1}$B$_{1}$	   & & & & & & -1.1104(1)      & & & & -1.1084(2)	& &  & & -1.1081[n=30]   &  & &  -1.1081	   \\
\\
 8       &  &  12	 & & & & $^{1}$A$_{1}$	   & & & & & & -0.9376(1)      & & & & -0.9329(5)	& &  &  & -0.9327[n=35]   &  & &   -0.9328	  \\
\\
 12      &  &  12	 & & & & $^{1}$A$_{1}$	   & & & & & & -0.8557(1)      & & & & -0.8507(6)	& &  &  & -0.8509[n=30]   &  & & -0.8512	  \\
\\
 4       &  &  14	 & & & & $^{1}$B$_{1}$	   & & & & & & -0.9863(1)      & & & & -0.9840(1)	& & &  & -0.9840[n=20]    &  & & -0.9840	  \\
\\
 8       &  &  14	 & & & & $^{1}$B$_{1}$	   & & & & & & -0.7461(1)      & & & & -0.7417(8)	& & &  & -0.7417[n=40]    &  & &  -0.7418	  \\
\\
 12      &  &  14	 & & & & $^{1}$B$_{1}$	   & & & & & &  -0.6296(2)     & & & & -0.627(4)	& & &  & -0.6281[n=30]    &  & & -0.6282	  \\
\\
 4       &  &  16	 & & & & $^{1}$A$_{1}$	   & & & & & & -0.85140(6)     & & & & -0.85133(6)	& & &  & -0.85134[n=10]   &  & &  -0.85137	   \\
\\
 8       &  &  16	 & & & & $^{1}$A$_{1}$	   & & & & & & -0.5293(2)      & & & & -0.5291(2)	& & &  & -0.5293[n=10]    &  & &  -0.5293	      \\
\\
 12      &  &  16	 & & & & $^{1}$A$_{1}$	   & & & & & & -0.3741(2)      & & & & -0.3739(4)	& & &  & -0.3745[n=10]    &  & &  -0.3745	      \\
\\
\hline
\end{tabular}
\end{table*}

Within the FED approach, \cite{rayner-Hubbard-1D-FED2013,rayner-Hubbard-1D-FED2014}
we consider a set of spin and space group symmetry-broken HF determinants
$| {\cal{D}}^{i} \rangle$ ($i=1, \dots$, n).  Each of these
determinants $| {\cal{D}}^{i} \rangle$ is a convenient
mean field (intrinsic) trial configuration.
We restore the symmetries of the 2D Hubbard Hamiltonian Eq.(\ref{HAM-hubbard1D}).
resorting to projection techniques. \cite{rs,Carlo-review,Schmid-Gruemmer-1984}
 Let us denote
$R(\Omega)$ and  $\hat{R}(g)$ as the
symmetry operations associated with the spin and
space groups, respectively, parametrized in terms of the
 Euler angles $\Omega$=$(\alpha,\beta,\gamma)$
and the label g for the corresponding point group
operations.
One then uses the degeneracy  of the Goldstone states
$| {\cal{D}}^{i} (\Omega,g)\rangle = \hat{R}(\Omega) \hat{R}(g) | {\cal{D}}^{i} \rangle$
to recover the desired global gauge
symmetries  by means of a MR FED wave function
of the form

\begin{eqnarray} \label{FED-state-general}
| \phi_{K}^{ \Theta} \rangle =
\sum_{K^{'}}
\sum_{i=1}^{n}
f_{K^{'} }^{i \Theta}
\hat{P}_{K K^{'}}^{\Theta}
| {\cal{D}}^{i} \rangle
\end{eqnarray}
which expands a given ground state $| \phi_{K}^{ \Theta} \rangle$, with well
defined spin and space group  quantum numbers $\Theta$, in terms of n nonorthogonal
 symmetry-projected basis states.
 The operator $\hat{P}_{K K^{'}}^{\Theta}$
 takes the form

\begin{eqnarray} \label{projection-operator-P}
\hat{P}_{K K^{'}}^{\Theta}= \frac{h}{L} \sum_{m}^{L} {\Gamma}_{K K^{'}}^{\Theta *}(m)
\hat{R}(m)
\end{eqnarray}
where the sum runs  over all the symmetry transformations realized by
$\hat{R}(\Omega,g)$.

The quantity ${\Gamma}_{K K^{'}}^{\Theta }(m)$ represents
the character of the  irreducible representation
\cite{Edmonds,Tomita-1,Lanczos-Fano}
while h and L are the dimension of the irreducible representation and
the order of the
corresponding symmetry group, respectively. In the case of the continuous SU(2)
spin-rotational symmetry, the sum should be understood as a group
integration  with the appropriate
measure. \cite{Edmonds}

\begin{table*}
\label{Table2}
\caption{Ground state energies (in t units)
obtained with the CPQMC approach based on
 SR symmetry-projected UHF (SR-UHF-CPQMC) and GHF (SR-GHF-CPQMC)  states
 \cite{PaperwithShiwei}
are compared
with those obtained with MR  UHF-FED and GHF-FED calculations, based on n
transformations, in the case of
a 4 $\times$ 4 lattice with 12 and
14 electrons at U=4t and 8t.
For each configuration
the corresponding set of symmetry quantum numbers $\Theta$
[in all cases ${\bf{k}}$=(0,0)]
is included in the table.
Exact diagonalization (ED) results
\cite{Shiwei-QMC-Symmetry,PaperwithShiwei}
 are listed in the last column.
 For more details, see the main text.
}
\begin{tabular}{cccccccccccccccccccccccccccc}
\hline
\\
 $U/t$    &  &$N_{e}$  & & & & $\Theta$          & & & & & &   UHF-CPQMC    & & & &  UHF-FED[n]	        & & &  &  GHF-CPQMC	 & & &  & GHF-FED[n]	    &  & &  ED \\
\\
\hline
\\
 4       &  &  12	 & & & & $^{1}$B$_{1}$	   & & & & & &  -17.7327(8)    & & & & -17.7293[n=130]  & &  & & -17.7301(1)	 & &  & & -17.7296[n=30]    &  & &  -17.7296		\\
\\
 8       &  &  12	 & & & & $^{1}$A$_{1}$	   & & & & & &  -14.914(3)     & & & & -14.9227[n=160]  & &  & & -14.920(1)	 & &  & & -14.9232[n=35]    &  & &  -14.925         \\
\\
 4       &  &  14	 & & & & $^{1}$B$_{1}$	   & & & & & &  -15.7482(5)    & & & & -15.7422[n=50]   & & &  & -15.7455(2)	 & & &  & -15.7440 [n=20]    &  & & -15.7446	       \\
\\
 8       &  &  14	 & & & & $^{1}$B$_{1}$	   & & & & & &  -11.872(3)     & & & & -11.8665[n=130]  & & &  & -11.847(3)	 & & &  & -11.8672[n=40]     &  & &  -11.8688          \\
\\
\hline
\end{tabular}
\end{table*}

The linear momenta
$k_{x}=\frac{2 \pi}{N_{x}}$ ${\xi}_{x}$
and $k_{y}=\frac{2 \pi}{N_{y}}$ ${\xi}_{y}$
are given in terms of the quantum numbers
${\xi}_{x}$ and ${\xi}_{y}$
which take the values allowed inside the Brillouin zone. \cite{Ashcroft-Mermin-book}
Obviously, since we consider the full space group, for certain
high symmetry momenta, additional parities $b_{x}$, $b_{y}$ and $b_{xy}$
under x, y and x-y reflections are needed.
For example, in the case of a square lattice
we refer to states
with
$\Theta=(0,0,0,+1,+1,+1)$ and $\Theta=(0,0,0,+1,+1,-1)$
as
$^{1}$A$_{1}$ and $^{1}$B$_{1}$ configurations, respectively, with ${\bf{k}}$=(0,0).
For the sake of brevity, we refer to these states in what follows
as $^{1}$A$_{1}$ and/or $^{1}$B$_{1}$
configurations . However, the reader should keep in mind that
we always use the full set of quantum numbers required to characterize a given
FED state.

In this study, we have considered two types
\cite{StuberPaldus,HFclassification}
 of intrinsic Slater determinants $| {\cal{D}}^{i} \rangle$
in the  expansion Eq.(\ref{FED-state-general}), i.e.,
unrestricted  (UHF) and generalized (GHF)  Hartree-Fock :

\begin{itemize}

\item UHF states preserve $\hat{S}_{z}$-symmetry
while possibly breaking all others. They preserve
N$_{e,\uparrow}$ and N$_{e,\downarrow}$ electron number.

\item GHF states break all Hubbard 2D Hamiltonian symmetries
and can only be characterized by N$_{e}$, the total number
of electrons.

\end{itemize}

In those cases where calculations are performed in terms
of symmetry-projected UHF configurations, the integrals in both Euler angles $\alpha$ and
$\gamma$, associated with the spin-projection operator,
become trivial and can be carried out analytically.
Calculations in terms of  symmetry-projected GHF states
are more elaborate as they necessitate numerical integrations over a three-dimensional
$(\alpha,\beta,\gamma)$-grid.
To indicate the type of Slater determinants used, we refer to the corresponding
VAP calculations as UHF-FED and GHF-FED, respectively. Obviously, for the same number
n of transformations, a GHF-FED ansatz is computationally more demanding but also
accounts for more correlations than the UHF-FED one,
because of its larger variational flexibility.
However, given the fact that
UHF-FED calculations are roughly two orders of magnitude less
computationally demanding than GHF-FED, in this study we have
also resorted to the former. This alleviates our numerical
effort and enable us to reach  larger lattices and/or a larger
number n of basis states in Eq.(\ref{FED-state-general}).

Regardless of the UHF or GHF symmetry-broken character
of the $| {\cal{D}}^{i} \rangle$ states used, the MR FED wave function $| \phi_{K}^{\Theta} \rangle$
is determined applying the variational principle  to the projected energy
\begin{eqnarray} \label{ojo-ojo}
E^{ \Theta} =
\frac{
f^{\Theta \dagger}
{\cal{H}}^{ \Theta}
f^{ \Theta}
}
{
f^{ \Theta \dagger}
{\cal{N}}^{ \Theta}
f^{ \Theta}
}
\end{eqnarray}
written in terms of the Hamiltonian and norm overlaps
\begin{eqnarray} \label{HNKernels-GHF-FED}
{\cal{H}}_{i K,j K^{'}}^{ \Theta} &=&
\langle {\cal{D}}^{i} | \hat{H} \hat{P}_{K K^{'}}^{\Theta}  | {\cal{D}}^{j} \rangle
\nonumber\\
{\cal{N}}_{i K,j K^{'}}^{ \Theta} &=&
\langle {\cal{D}}^{i} |  \hat{P}_{K K^{'}}^{\Theta}  | {\cal{D}}^{j} \rangle
\end{eqnarray}
between all the symmetry-projected configurations
in
Eq.(\ref{FED-state-general}). All the
matrix elements
$\langle {\cal{D}}^{i} | \hat{H} \hat{R}(\Omega,g)  | {\cal{D}}^{j} \rangle$
and $\langle {\cal{D}}^{i} | \hat{R}(\Omega,g)  | {\cal{D}}^{j} \rangle$
needed to compute the kernels Eq.(\ref{HNKernels-GHF-FED})
can be found with the help of the extended Wick theorem. \cite{Blaizot-Ripka}
For the mixing coefficients $f^{i \Theta}$
we obtain a resonon-like \cite{Gutzwiller_method} eigenvalue equation
\begin{eqnarray} \label{HW-1}
\left({\cal{H}}^{ \Theta}
- E^{\Theta}
{\cal{N}}^{ \Theta}
\right)
f^{\Theta} = 0
\end{eqnarray}
with the constraint
$f^{ \Theta \dagger} {\cal{N}}^{ \Theta} f^{ \Theta} = 1$ ensuring
the normalization of the wave function. The energy Eq.(\ref{ojo-ojo}) is varied only
with respect to the last added  determinant
$| {\cal{D}}^{n} \rangle$ keeping all the other transformations
${\cal{D}}^{i}$ (i=1, $\dots$, $n-1$), obtained in
previous VAP calculations, fixed. \cite{rayner-Hubbard-1D-FED2013,Carlo-review,Carlos-Rayner-Gustavo-FED-molecules}
We have parametrized the variation with respect to each of the
 transformations  ${\cal{D}}^{i}$
in terms of the Thouless theorem. \cite{rs} Such a parametrization has already been
shown to be a useful tool in nuclear structure \cite{Rayner-Robledo-fission-2014,Carlo-review,Rayner-Carlo-CM-1,Rayner-Carlo-CM-2}
and condensed matter \cite{Carlos-Hubbard-1D,Rayner-2D-Hubbard-PRB-2012,non-unitary-paper-Carlos,rayner-Hubbard-1D-FED2013,rayner-Hubbard-1D-FED2014}
physics but also in quantum
chemistry. \cite{Carlos-Rayner-Gustavo-VAMPIR-molecules,Carlos-Rayner-Gustavo-FED-molecules,Laimis-paper}

All the results discussed  in this paper have been obtained
with an in-house code where the optimization is handled with
a limited-memory quasi-Newton method. \cite{quasi-Newton}
In a previous study, \cite{rayner-Hubbard-1D-FED2013} we have discussed the computational performance
of the FED scheme in the case of 1D systems. It has been shown that
its speedup grows linearly with the number of processors
used in the calculations. On the other hand, for a fixed number
of processors, an efficient implementation
scales linearly with the number of symmetry-projected
configurations $\hat{P}^{\Theta} | {\cal{D}}^{i} \rangle$ used.  A
typical outcome of our calculations is shown in Fig. \ref{Fig1}
where we have plotted the UHF-FED speedup for a half-filled
12 $\times$ 12 lattice at U=4t. Note that an efficiency of
almost 100 $\%$ in the parallelization is observed
even when the calculations are run on tens of thousands of processing cores.

\section{Discussion of results}
\label{results}

In this section, we discuss the results of our benchmark
calculations. First, in Sec. \ref{results-gs-corr-energies}, we
 compare the predicted ground state and correlation energies
 for half-filled and doped 4 $\times$ 4, 6 $\times$ 6, 8 $\times$ 8, and
 10 $\times$ 10 lattices with those obtained using other theoretical approaches. Most of the calculations have been carried out
 at U=2t, 4t, 8t, and 12t.
 We also discuss the
dependence of the predicted correlation energies on the
number of basis states used in the corresponding FED expansions and
the structure of the intrinsic determinants resulting from our
VAP procedure. Next, in Sec. \ref{Corre-functions}, we consider
the momentum distributions, SSCFs, CCCFs and PCFs. Results are
presented for a small 4 $\times$ 4 lattice with N$_{e}$=14 electrons but also
for a larger 16 $\times$ 4 one with N$_{e}$=56 electrons.

\subsection{Ground state energies, correlation energies and
structural
defects}
\label{results-gs-corr-energies}

In Fig. \ref{Fig2}, we have plotted (red diamonds) the relative energy
errors provided by the GHF-FED approach based on n=10 transformations.
They are compared with SR  results (blue diamonds) as well as with those
obtained within a  VMC scheme based on a CPS-Pfaffian ansatz
 \cite{Neuscamman-2012}
 (black diamonds). Results
are shown  at half-filling for U=2t, 4t, 8t, 12t and 16t. In all cases
the ground states are characterized by $^{1}$A$_{1}$ symmetry. As can
 be seen, the  GHF-FED approach
outperforms the CPS-Pfaffian-VMC one and is exact \cite{Lanczos-Fano,Neuscamman-2012,Shiwei-QMC-Symmetry,Dagotto-Moreo-1992}
to all the
considered figures. Note that, for this small system, a SR approach  already
 provides relative errors smaller than 0.04 $\%$.

%
%
\begin{figure}
\includegraphics[width=0.45\textwidth]{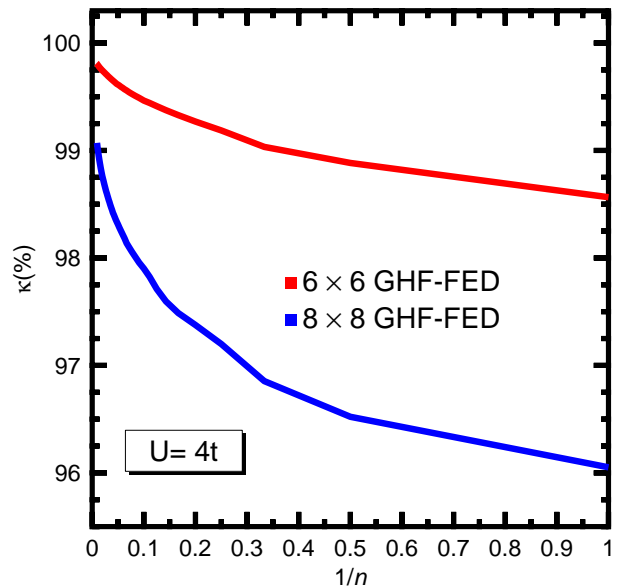}
\caption{(Color online)  The percentage of correlation energy $\kappa$
[Eq.(\ref{formulaCE})]
obtained with the
GHF-FED approximation is plotted as a function of the inverse of the number n of
transformations for the half-filled
6 $\times$ 6 (red curve) and 8 $\times$ 8 (blue curve) lattices. Results
are shown for U=4t.
}
\label{Fig3}
\end{figure}
%
%

The auxiliary-field QMC approximation is an important
tool for studying correlated electronic systems.
\cite{Stripes-Shiwei}
The relevance of symmetries within this
framework has already been discussed in the
literature. \cite{Shiwei-QMC-Symmetry}
In Table 1, we compare ground state energies per site provided by our GHF-FED scheme for the 4 $\times$ 4 lattice at different doping fractions $x$
and onsite interactions of U=4t, 8t, and 12t
with those
obtained within the constrained-path (CPQMC) and release-constraint (RCQMC) QMC approximations. Both
the CPQMC and RCQMC calculations were based on multideterminantal trial wave functions
with  symmetries obtained
in the spirit of a small complete active-space self-consistent field (CASSCF) calculation. \cite{Shiwei-QMC-Symmetry}
For each configuration,
the corresponding set of symmetry quantum numbers $\Theta$ is also given in the table. It is satisfying to
observe
that for the considered number n of symmetry-projected configurations, the GHF-FED energies are
slightly
more accurate than
 the CPQMC ones
 and reach the accuracy of the  RCQMC results. The comparison with
ED calculations, in the
last column of the table, indicates that for this small lattice both the GHF-FED and RCQMC energies
can be
considered exact
\cite{Shiwei-QMC-Symmetry,Dagotto-Moreo-1992}
for  practical purposes.

%
%
\begin{figure*}
\includegraphics[width=0.490\textwidth]{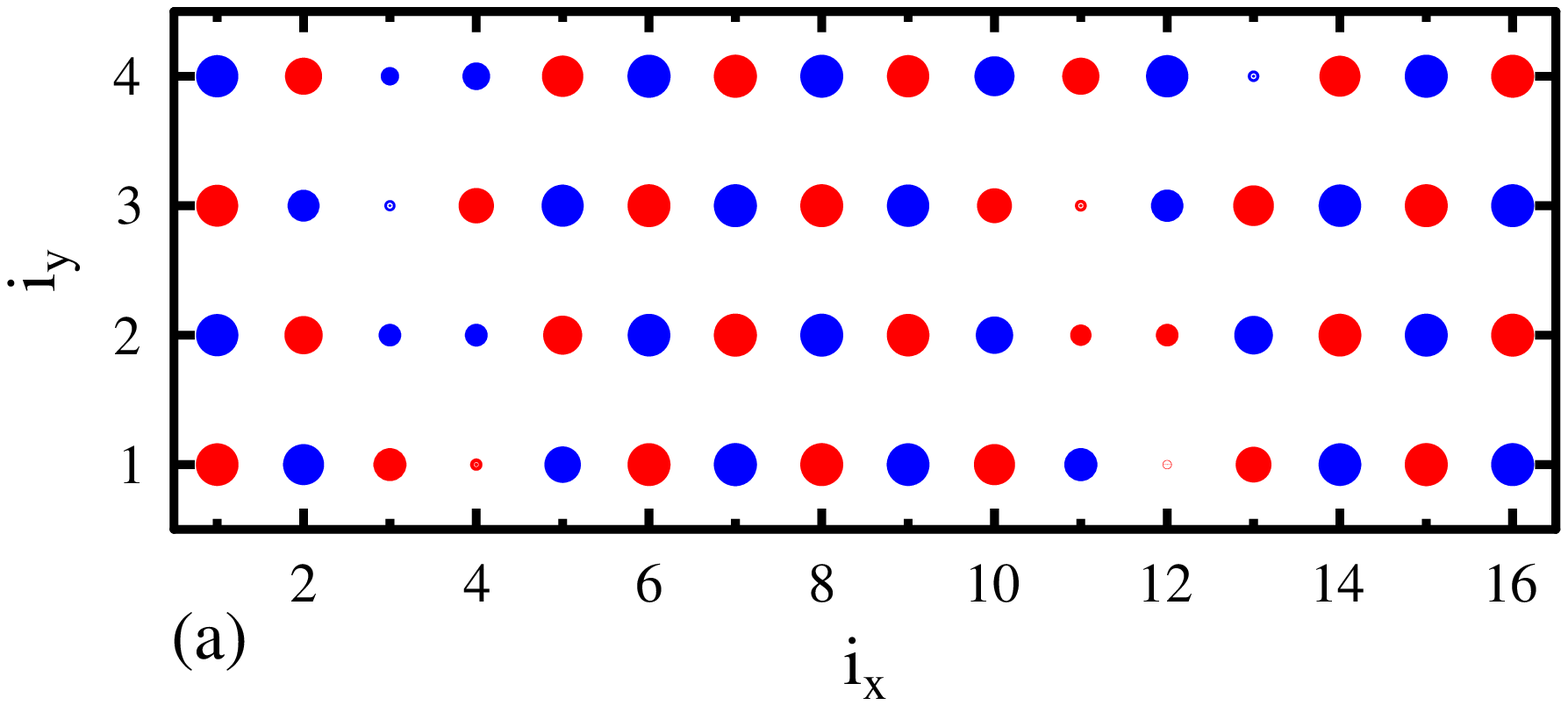}
\includegraphics[width=0.490\textwidth]{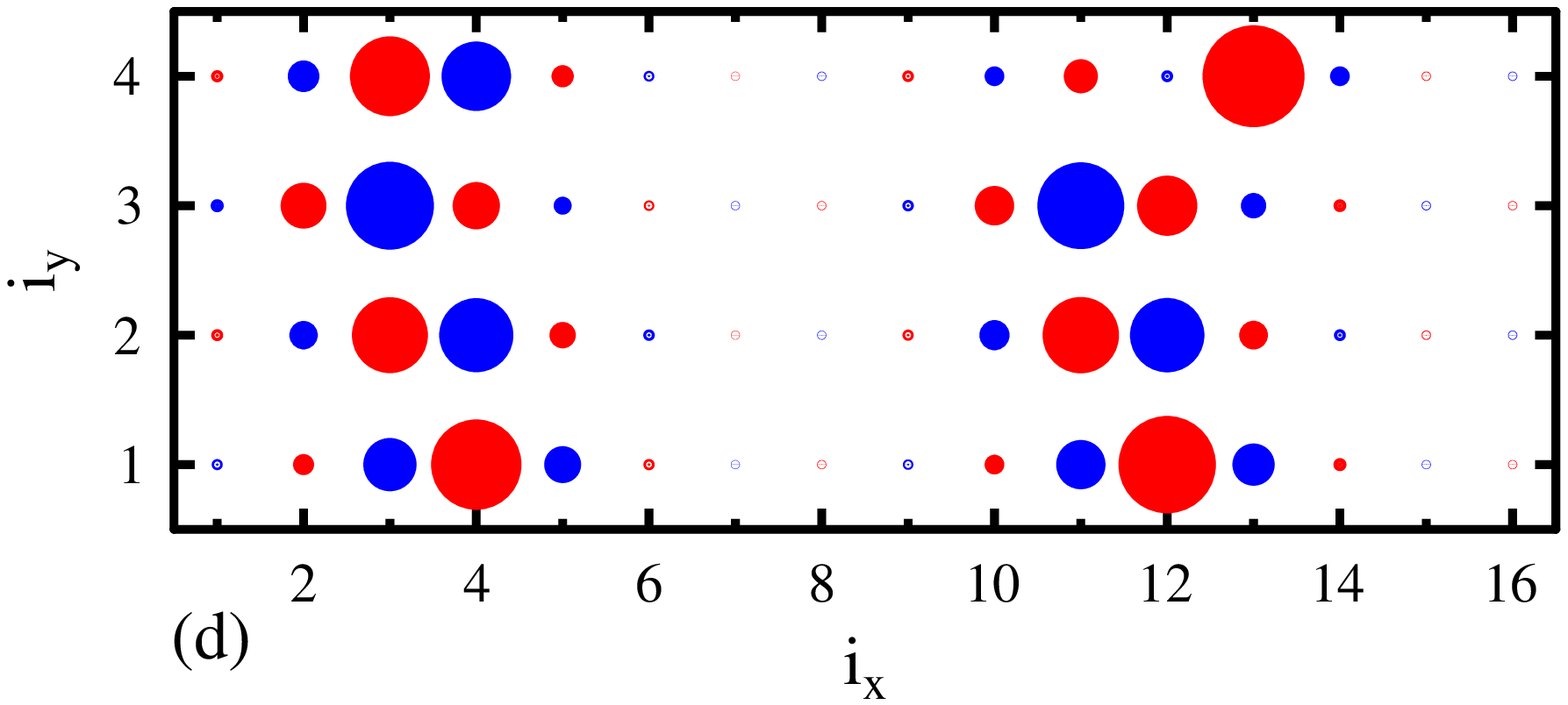} \\
\includegraphics[width=0.490\textwidth]{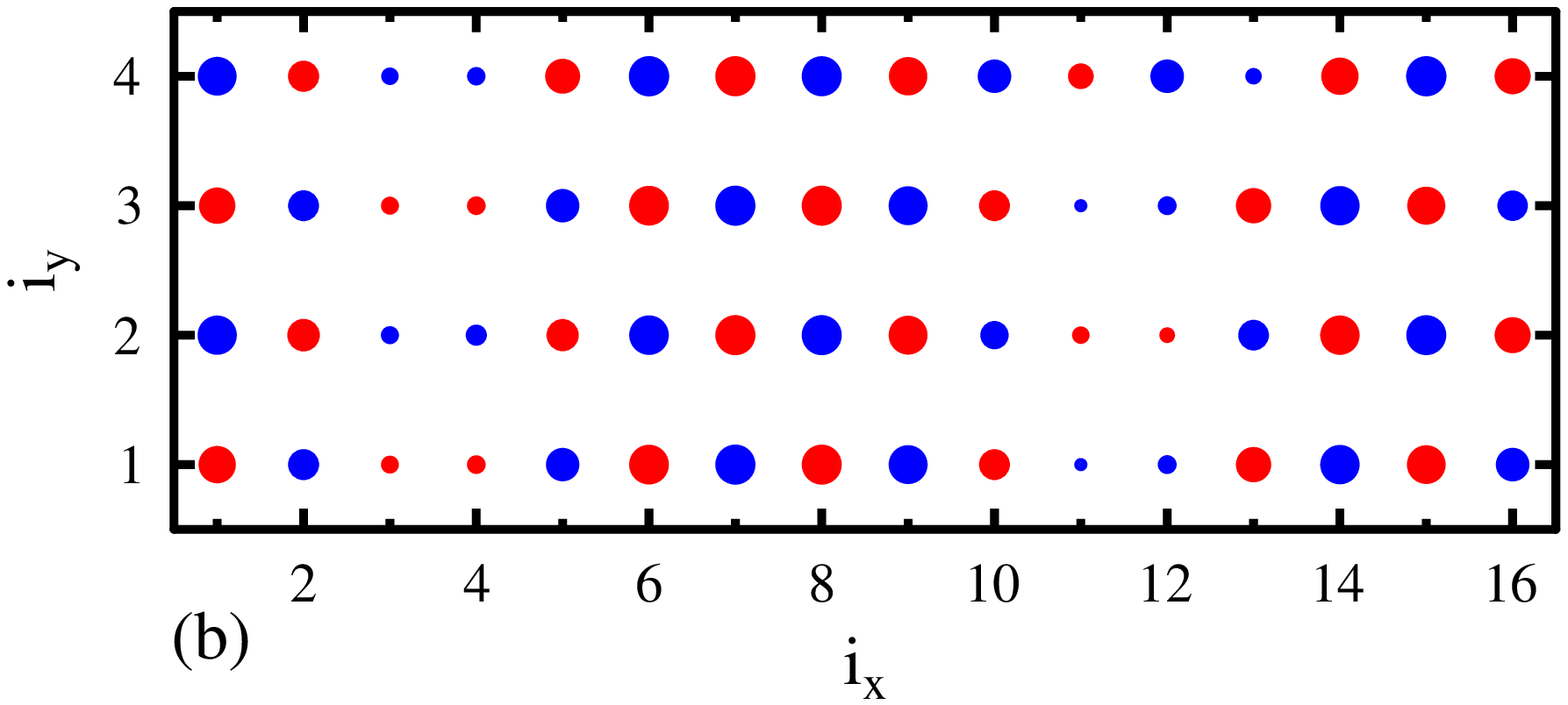}
\includegraphics[width=0.490\textwidth]{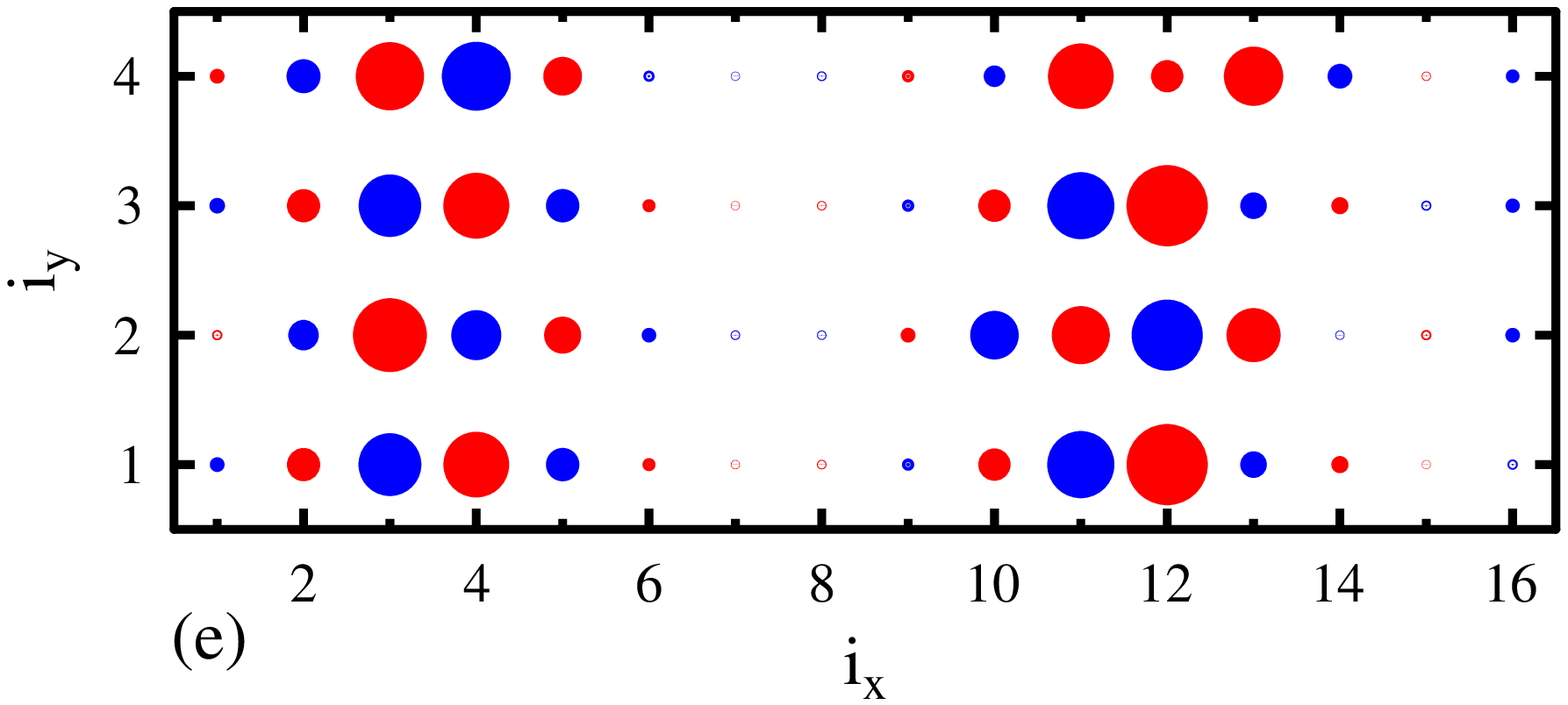} \\
\includegraphics[width=0.490\textwidth]{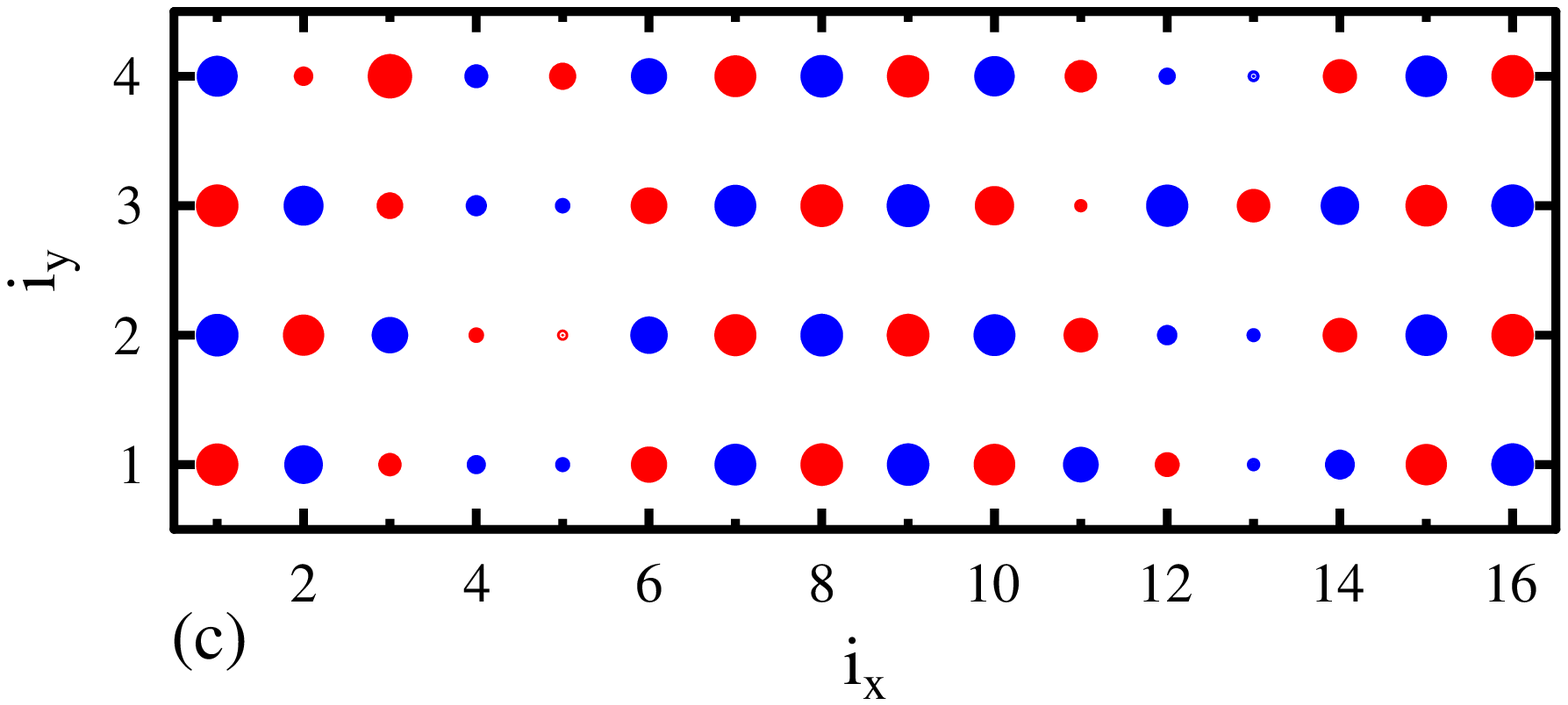}
\includegraphics[width=0.490\textwidth]{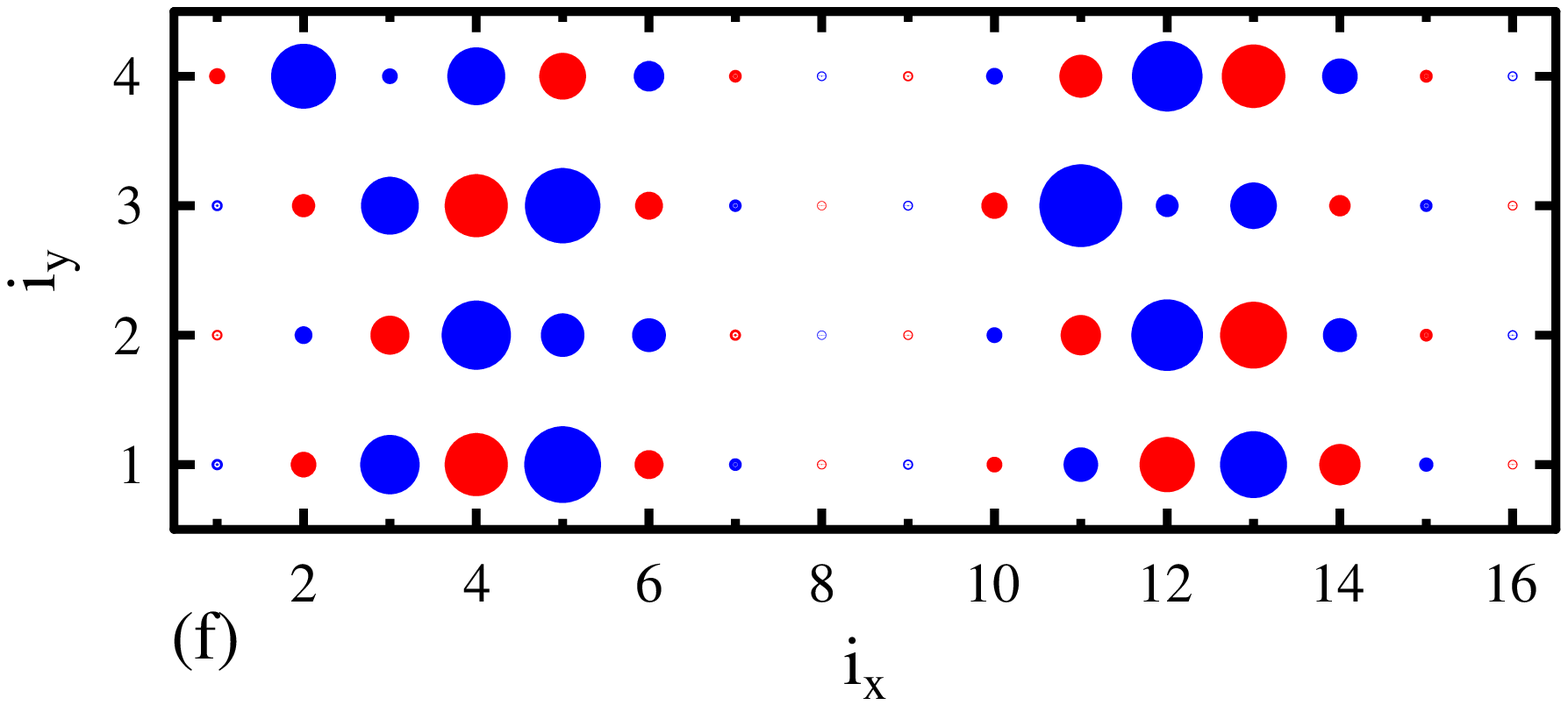}\\
\caption{(Color online) The spin $\xi^{r}({\bf{i}})$ [Eq.(\ref{SD-FED})]
and
charge  $\xi_{ch}^{r}({\bf{i}})$ [Eq.(\ref{CD-FED})] densities
obtained with some typical intrinsic  determinants
resulting from the UHF-FED variational scheme
are depicted in panels (a) to (c) and (d) to (f), respectively.
Results are shown, for the 16 $\times$ 4 lattice
with $N_{e}$=56 electrons at U=8t. The size of the circles is proportional to the value of the densities
at a given lattice site ${\bf{i}}$. Blue (red) circles refer to spin
up (down) in panels (a) to (c) and positve (negative) charge in panels (d) to (f). For details, see the main
text.
}
\label{Fig4}
\end{figure*}
%
%

We have recently explored the role of
SR symmetry-projected  trial states within the CPQMC
framework. \cite{PaperwithShiwei}
It has been shown, through a hierarchy of
calculations based  on
symmetry-projected trial states, that they
increase the energy
accuracy and decrease the statistical variance as more
symmetries are broken and restored.
The energies
obtained within the CPQMC approach based on
 SR symmetry-projected UHF  and GHF states (here, we use
the acronyms SR-UHF-CPQMC and SR-GHF-CPQMC, \cite{ACRONY} respectively)
are compared in Table II
with those obtained via MR  UHF-FED and GHF-FED calculations
in
the case of
a 4 $\times$ 4 lattice with 12 and
14 electrons at U=4t and 8t. One observes a good agreement
between all these approximations that compare well with
the ED results
\cite{Shiwei-QMC-Symmetry,PaperwithShiwei}
listed in the last column of the table.
Note also that the energies  in Tables I and II vastly  improve
those obtained in our previous study
\cite{Rayner-2D-Hubbard-PRB-2012}
of the 4 $\times$ 4  Hubbard
model as a result of the
(more correlated)
MR character of the ansatz of Eq.(\ref{FED-state-general})
which also incorporates restoration of the space group symmetry.

%
%
\begin{figure*}
\includegraphics[width=0.490\textwidth]{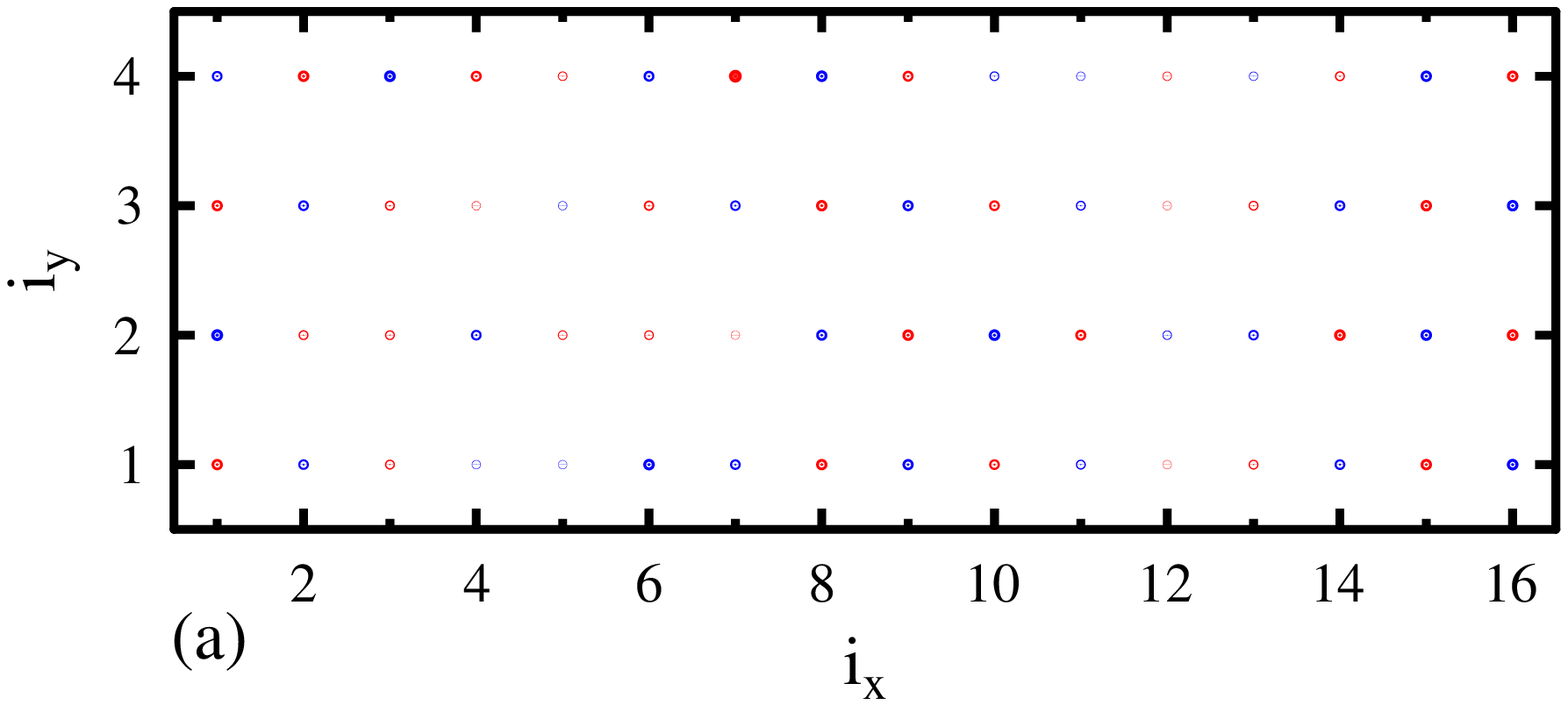}
\includegraphics[width=0.490\textwidth]{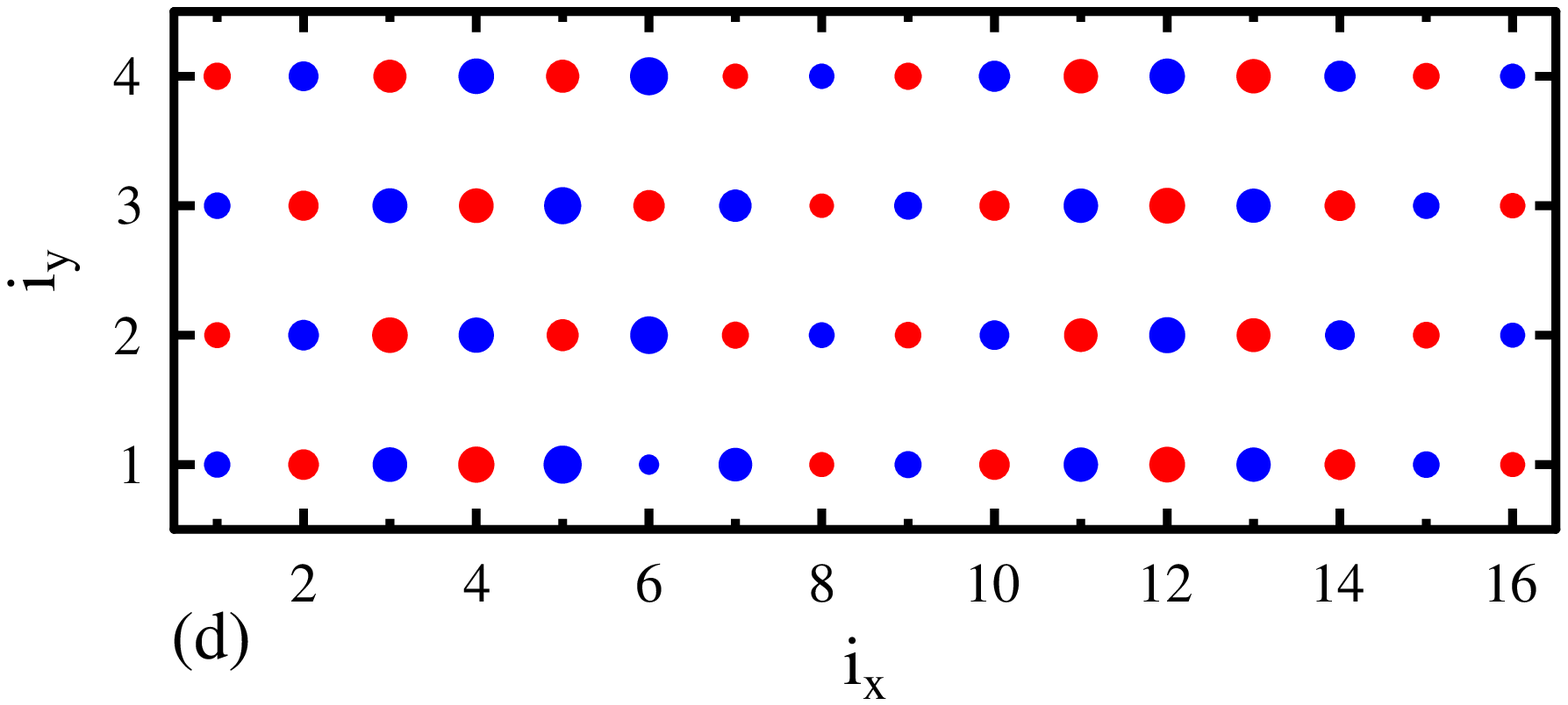} \\
\includegraphics[width=0.490\textwidth]{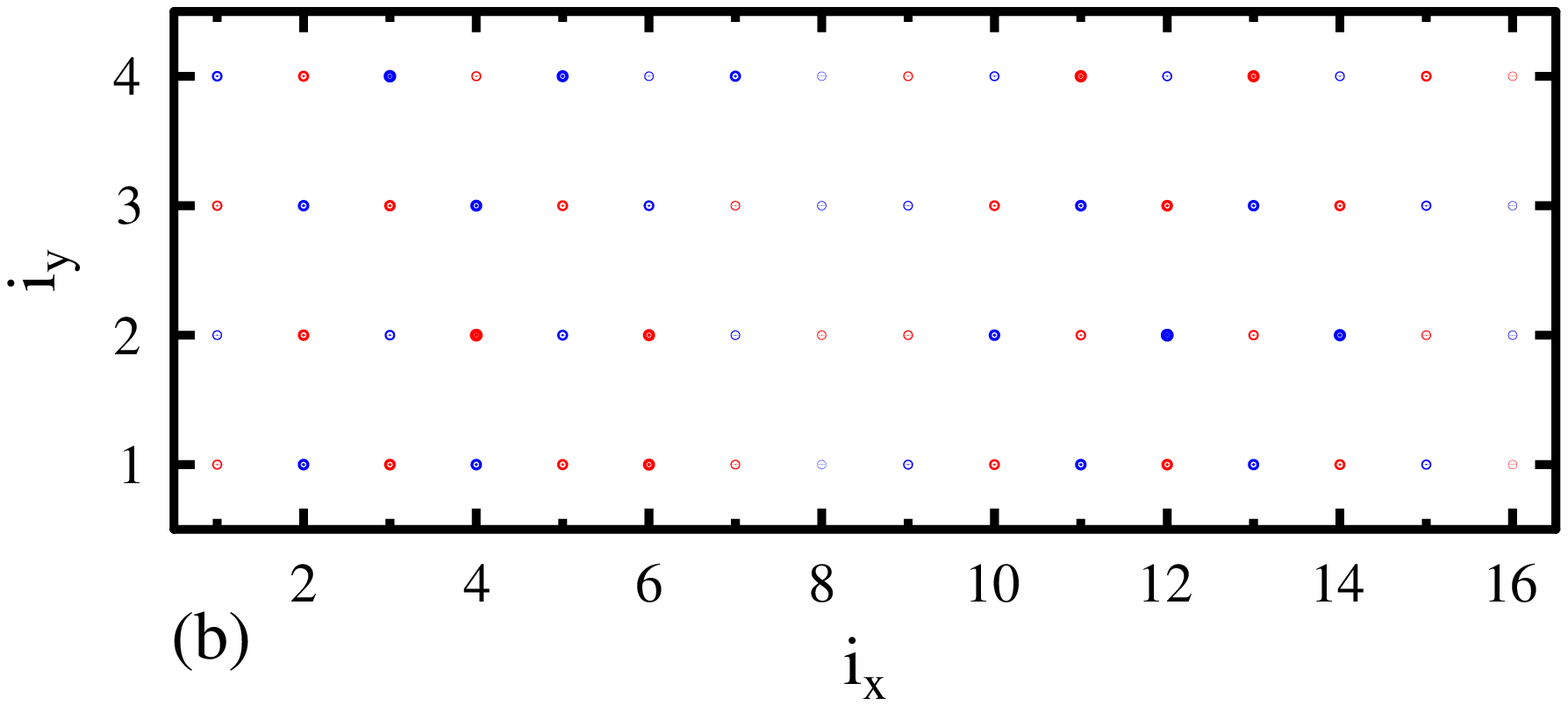}
\includegraphics[width=0.490\textwidth]{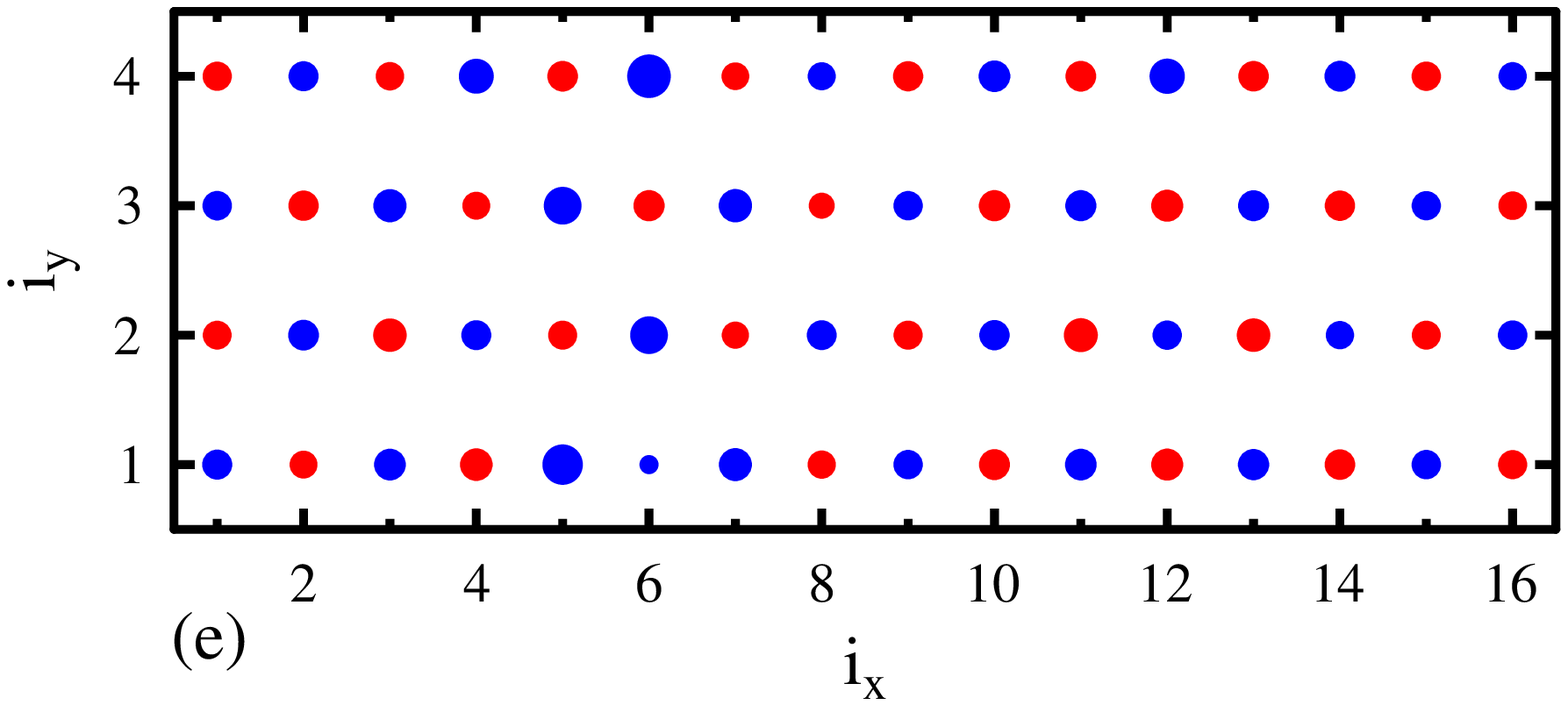} \\
\includegraphics[width=0.490\textwidth]{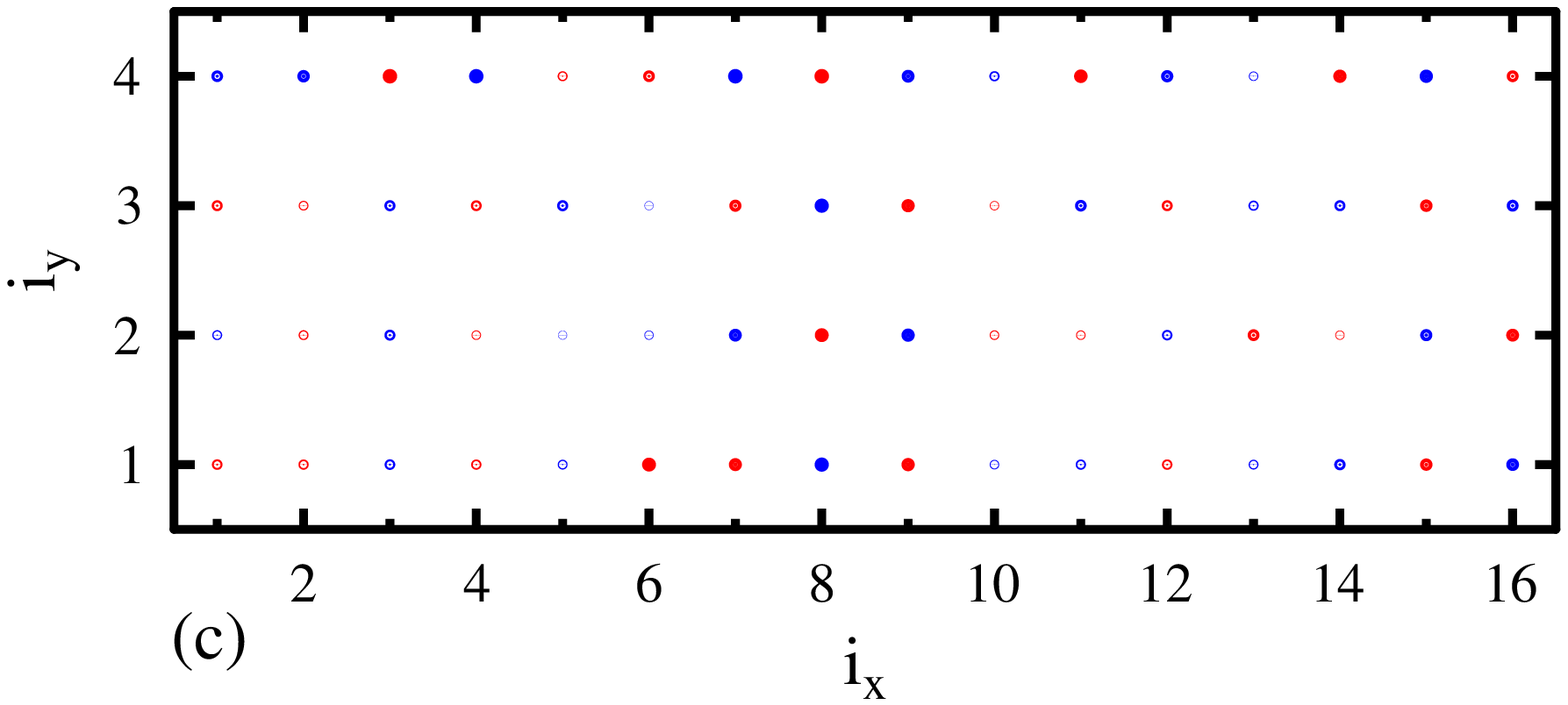}
\includegraphics[width=0.490\textwidth]{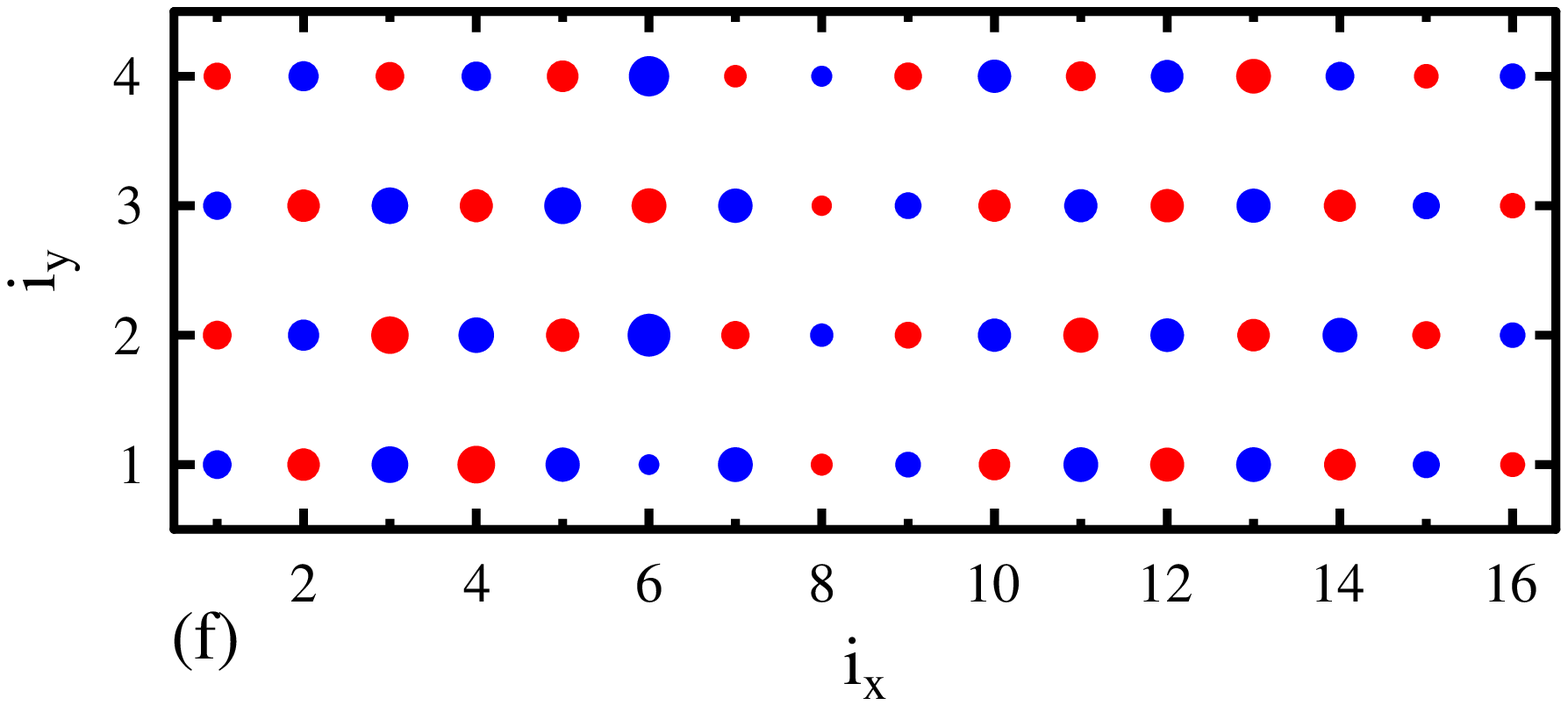}\\
\caption{(Color online) The same as Fig. \ref{Fig4} but for U=2t.
}
\label{Fig5}
\end{figure*}
%
%

The previous results illustrate that for half-filled and doped
 lattices up to 16 sites, the FED scheme
provides essentially exact ground state energies. The question
naturally arises as to whether reasonably correlated wave functions
can also be obtained for larger 2D systems. The percentage of correlation energies

\begin{eqnarray} \label{formulaCE}
\kappa_{GHF-FED} = \frac{E_{RHF}-E_{GHF-FED}}{E_{RHF}-E_{EXACT}} \times 100 \%
\end{eqnarray}
obtained with the
GHF-FED approximation is plotted
in  Fig. \ref{Fig3}
as a function of the inverse of the number n of
transformations for the half-filled
6 $\times$ 6 (red curve) and 8 $\times$ 8 (blue curve) lattices
at U=4t. The corresponding ground states are characterized by
 the $^{1}$B$_{1}$ and $^{1}$A$_{1}$ symmetries, respectively.
In Eq.(\ref{formulaCE}), $E_{RHF}$
is the energy of the standard restricted
HF (RHF) solution, \cite{StuberPaldus,HFclassification} that preserves all the
symmetries of the 2D Hubbard Hamiltonian. For the exact ground state energies of the
6 $\times$ 6 and 8 $\times$ 8 systems, we have used the
estimates
 -30.89(1)t and -55.09(3)t, respectively. \cite{Shiwei-private-6by6,aux-QMC-Imada}

The first noticeable feature from Fig. \ref{Fig3} is that even a SR
calculation recovers a large portion
of the correlation energy ($\kappa_{GHF-FED}$= 98.57 $\%$ and 96.05 $\%$
for the 6 $\times$ 6 and 8 $\times$ 8 lattices, respectively).
These values already represent a vast improvement with respect to the
standard UHF ones
(79 $\%$ and 77 $\%$). The correlation energies
increase smoothly with the number
of  symmetry-projected GHF basis states included in the FED expansion.
It is also apparent from the figure that with increasing lattice size,
we need to increase the number n of transformations to keep and/or
improve the quality of our MR wave functions.
For example, while n=10 transformations provide an essentially exact ground state
for the half-filled 4 $\times$ 4 lattice (see, Table I), the corresponding energies (i.e., -30.8316t and
-54.7157t) lead us to $\kappa_{GHF-FED}$=99.46 $\%$ and 97.90 $\%$
for the  6 $\times$ 6 and 8 $\times$ 8  ones. On the other hand, a further increase  of
the number of symmetry-projected GHF configurations up to n=120 and n=108 provides
the ground state energies -30.8695t and -54.9242t ($\kappa_{GHF-FED}$= 99.81 $\%$ and 99.07 $\%$).
Note that, in the case of the 6 $\times$ 6 lattice, our results also improve significantly the
energy (i.e., -30.5766t) reported in our previous study. \cite{Rayner-2D-Hubbard-PRB-2012}
A similar behavior is observed away from half-filling. For example, for a
 6 $\times$ 6 lattice with N$_{e}$=24 electrons ($^{1}$B$_{1}$ symmetry)
at U=4t we have
obtained the energies per site of -1.17884t and -1.18445t with n=10 and n=180 symmetry-projected GHF
configurations. The energies provided by the CPQMC and RCQMC approximations based on
CASSCF
multideterminantal
trial wave functions with symmetries are -1.18625(3)t and -1.18525(4)t, respectively. \cite{Shiwei-QMC-Symmetry}

The previous examples, and the results already obtained for 1D
lattices, \cite{rayner-Hubbard-1D-FED2013,rayner-Hubbard-1D-FED2014} reveal the inner
workings of the FED approach: it is a MR VAP strategy to build
reasonably correlated ground states, with well defined symmetry quantum numbers $\Theta$, in
low-dimensional
electronic systems. In fact, it represents  a constructive
approximation in which, regardless
of the dimensionality of the considered lattice, the quality
of the  ansatz Eq.(\ref{FED-state-general}), can be systematically
improved by increasing the number  of nonorthogonal symmetry-projected basis states
through chains of VAP calculations. Note that the FED wave function
is a discretized form of the exact coherent state representation of a
fermion state
\cite{Perlemov}
and therefore becomes exact in the limit n $\rightarrow$ $\infty$. In practical
applications, however, one
is always limited to a finite
set of transformations whose precise number for obtaining a given accuracy
is difficult to assert beforehand because the
nature of the underlying quantum fluctuations varies for different doping fractions
$x$ and onsite repulsions (see, below). In the examples
discussed above, we have especially tailored the number of
symmetry-projected
basis states
so as to compare well with state-of-the-art ground state energies
available in the literature. However, the constructive nature of the FED ansatz also allows us to
specifically
tailor the number of  symmetry-projected basis states
to capture the main  trends in the physical properties
of interest (see, Sec. \ref{Corre-functions}).

%
%
\begin{figure}
\includegraphics[width=0.490\textwidth]{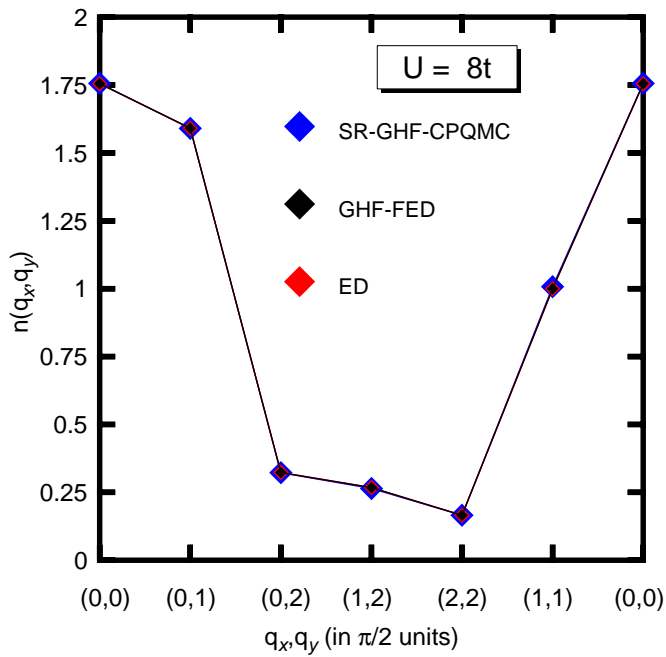}
\caption{(Color online) The momentum distribution [Eq.(\ref{nkdist})] provided
by GHF-FED calculations (black diamonds) , based on n=40 transformations (see Table I), is compared with that
 obtained within the SR-GHF-CPQMC approach
(blue diamonds)
as well as with ED results (red diamonds). \cite{PaperwithShiwei}
Results are shown for a 4 $\times$ 4 lattice with N$_{e}$=14 electrons at U=8t.
 For details, see the main text.
}
\label{Fig6}
\end{figure}
%
%

We have considered two order parameters, i.e., the
spin density (SD)

\begin{eqnarray} \label{SD-FED}
\xi^{r}({\bf{i}}) =
\langle {\cal{D}}^{r} | {\bf{S}}({\bf{i}}) | {\cal{D}}^{r} \rangle
\langle {\cal{D}}^{r} | {\bf{S}}({\bf{1}}) | {\cal{D}}^{r} \rangle
\end{eqnarray}
and the charge density (CD)

\begin{eqnarray} \label{CD-FED}
\xi_{ch}^{r}({\bf{i}}) =1 - \sum_{\sigma} \langle {\cal{D}}^{r} | \hat{n}_{{\bf{i}} \sigma} | {\cal{D}}^{r} \rangle
\end{eqnarray}
associated with the  symmetry-broken determinants $| {\cal{D}}^{r} \rangle$ resulting from
UHF-FED calculations for a  16 $\times$ 4 lattice with N$_{e}$=56 electrons
($\delta=1-x$=1/8)
at  U=2t, 4t, 8t and 12t. We have restricted ourselves to n=40
symmetry-projected basis states
which, as shown in the
next Sec. \ref{Corre-functions}, is enough to
capture the main features
of the considered correlation functions. The $^{1}$A$_{1}$ FED
wave functions have the energies
-82.1476t, -64.5369t, -44.4349t and -35.8096t, respectively. Obviously, these energies
can be further improved by increasing the number of UHF transformations. Thus, for
example, with n=200 we have obtained the value -65.1109t at U=4t. Note, that
with n=40 ours are, from the energetical point of view,  significantly better wave
functions than the ones obtained in a routine SR symmetry-projected
GHF calculation. \cite{Juillet} Among all the UHF  determinants $| {\cal{D}}^{r} \rangle$
used in the expansion Eq.(\ref{FED-state-general}) at U=8t and 2t, we have selected a typical subset
 to plot in Figs. \ref{Fig4} and \ref{Fig5}  the corresponding
SD  [panels (a)-(c)] and CD   [panels (d)-(f)] as functions of the lattice site ${\bf{i}}$.

It becomes apparent from Fig. \ref{Fig4} that the  Slater determinants
resulting from the UHF-FED VAP procedure contain structural defects
in the SD
at different lattice locations. In particular, they display vertical stripes
separated by $\Delta i_{x}$ = 1/$\delta$ = 8 sites
with
deviations (fluctuations)
from the pattern obtained with the UHF approximation. \cite{Xu-Chang-Walter-Shiwei-UHF-stripes}
 One also observes
that the charges delocalize within $\Delta i_{x}$ $\approx$ 1/2$\delta$ = 4 sites.
Similar results are obtained for U=12t. From a theoretical point of view, stripes
can be viewed as generic semiclassical instabilities in doped Mott-Hubbard insulators, \cite{Zaanen-Oles}
first pointed out by Tranquada et al. \cite{Tranquada} and subsequently studied
by several authors. \cite{Stripes-Shiwei,White-stripes,Hager-stripes}
The fluctuating stripes encode
one possible kind of
 basic unit of quantum fluctuations in the MR UHF-FED
wave functions.  We stress
that a doping x =7/8
is commensurate with the appearance of two stripes
in the  16 $\times$ 4 lattice. \cite{Stripes-Shiwei}
The comparison with Fig. \ref{Fig5} reveals that even though
defects are also present, the nature of the quantum
fluctuations is completely different in the weak interaction regime with the charges
spread all over the lattice. The same is also true for U=4t
although in this case the charges start to display a tendency to localize around particular lattice
sites. These results
already suggest a transition to a stripe regime for increasing U values,
as predicted within the auxiliary-field QMC framework. \cite{Stripes-Shiwei}
Furthermore,  since the space group
and spin
projection operators can only translate by one site and rotate
 defects in the intrinsic states $| {\cal{D}}^{r} \rangle$
but do not destroy them, one may expect, as shown to be the case in
Sec. \ref{Corre-functions}, that our MR symmetry-projected wave functions capture
such a transition and reflect it in the corresponding correlation functions.

A rich variety of defects is also
found in
other lattices at different doping fractions $x$ and/or
U values. We have also performed UHF-FED
calculations  for a 10 $\times$ 10 lattice
with N$_{e}$ = 92, 96 and 100  electrons at U=8t.
We have restricted ourselves to a sample of n=70 symmetry-projected basis states, which is
more than enough to obtain information about the basic units of quantum fluctuations
in the  intrinsic states $| {\cal{D}}^{r} \rangle$. The   energies
associated with the corresponding $^{1}$A$_{1}$ and $^{1}$B$_{1}$ states are
-60.6629t, -55.2559t and -50.8999t, respectively, which already improve the available
ResHF values. \cite{Tomita-2} For example, at
half-filling, we have found (neutral) T-shaped defects similar to those predicted
within the UHF-ResHF approximation. \cite{Tomita-2}
Due to the presence of several close lying solutions, which are a consequence of the
non linear character of the  UHF-FED and/or GHF-FED ans\"atze, a more detailed analysis
of the corresponding defects is left for future work. However we stress that similar
to the 1D case, \cite{Tomita-1,rayner-Hubbard-1D-FED2013,rayner-Hubbard-1D-FED2014} both the FED and the ResHF
VAP strategies provide intrinsic HF determinants whose defects encode information about
the basic units of quantum fluctuations in the considered 2D systems.

%
%
\begin{figure}
\includegraphics[width=0.420\textwidth]{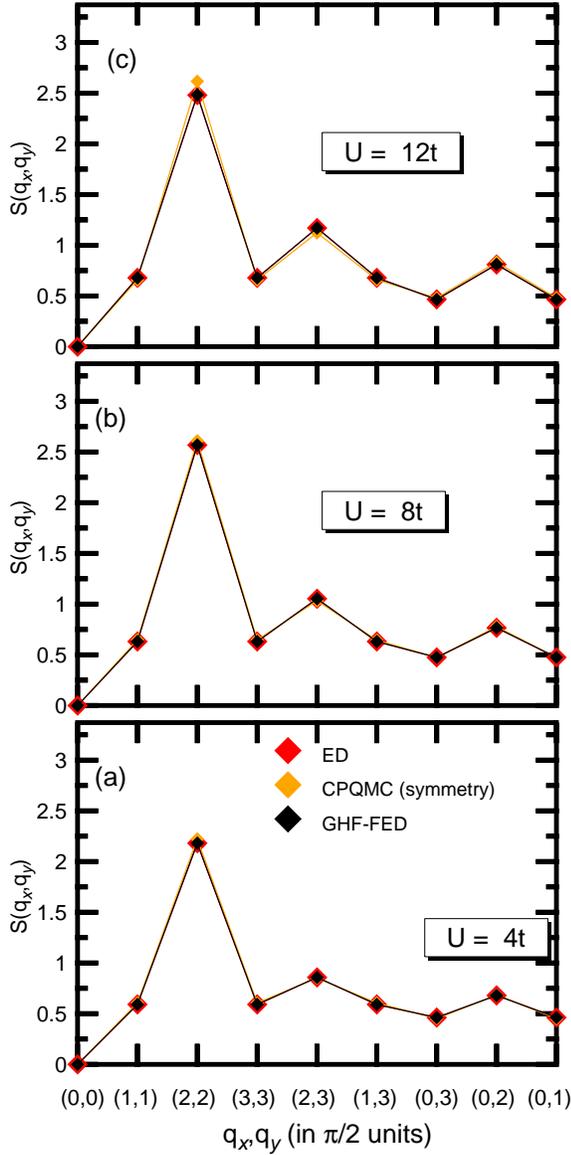}
\caption{(Color online) Fourier transforms of the GHF-FED ground-state spin-spin correlation functions in real space
[Eq.(\ref{ss-CF})]
for a 4 $\times$ 4 lattice with N$_{e}$= 14 electrons (black diamonds) are compared with CPQMC results
based on trial wave functions with symmetries  (orange diamonds). ED  values are depicted with red diamonds.
\cite{Shiwei-QMC-Symmetry}
Results are shown for
the onsite interactions 4t (a), 8t (b) and 12t (c). For more details, see the main text.
}
\label{Fig7}
\end{figure}
%
%

\subsection{Momentum distributions and correlation functions}
\label{Corre-functions}

In this section, we turn our attention to both momentum distributions
and correlation functions. It has been shown
within the auxiliary-field QMC framework,  \cite{PaperwithShiwei,Shiwei-QMC-Symmetry}
that trial states with
symmetries  are important to  account for SSCFs
and momentum distributions. Therefore, it is
interesting to study to which extent our MR wave functions, with well
defined quantum
numbers $\Theta$, can account for the main trends in those physical quantities. To this end, we first discuss our benchmark
calculations for a 4 $\times$ 4 lattice. The momentum distribution reads

\begin{eqnarray} \label{nkdist}
n^{\Theta}({\bf{q}}) = \sum_{\sigma}
\frac{
\langle \phi_{K}^{\Theta} | {\hat{n}}_{{\bf{q}} \sigma} | \phi_{K}^{\Theta} \rangle
}
{
\langle \phi_{K}^{\Theta}  | \phi_{K}^{\Theta} \rangle
}
\end{eqnarray}
where ${\hat{n}}_{{\bf{q}} \sigma}$ is the $\sigma$-occupation at wave vector ${\bf{q}}$.

In Fig. \ref{Fig6}, we have plotted   the ground state momentum distribution  computed with
the GHF-FED scheme  (black diamonds)  based on n=40 transformations (see Tables I and  II),  for a 4 $\times$ 4 lattice
with N$_{e}$=14 electrons at U=8t.
It is  compared with the
one obtained within the SR-GHF-CPQMC approach
(blue diamonds),
as well as with ED results (red diamonds). \cite{PaperwithShiwei}
We observe an excellent
agreement between ours, the SR-GHF-CPQMC momentum distribution, and  the one  resulting
from ED calculations.

%
%
\begin{figure}
\includegraphics[width=0.525\textwidth]{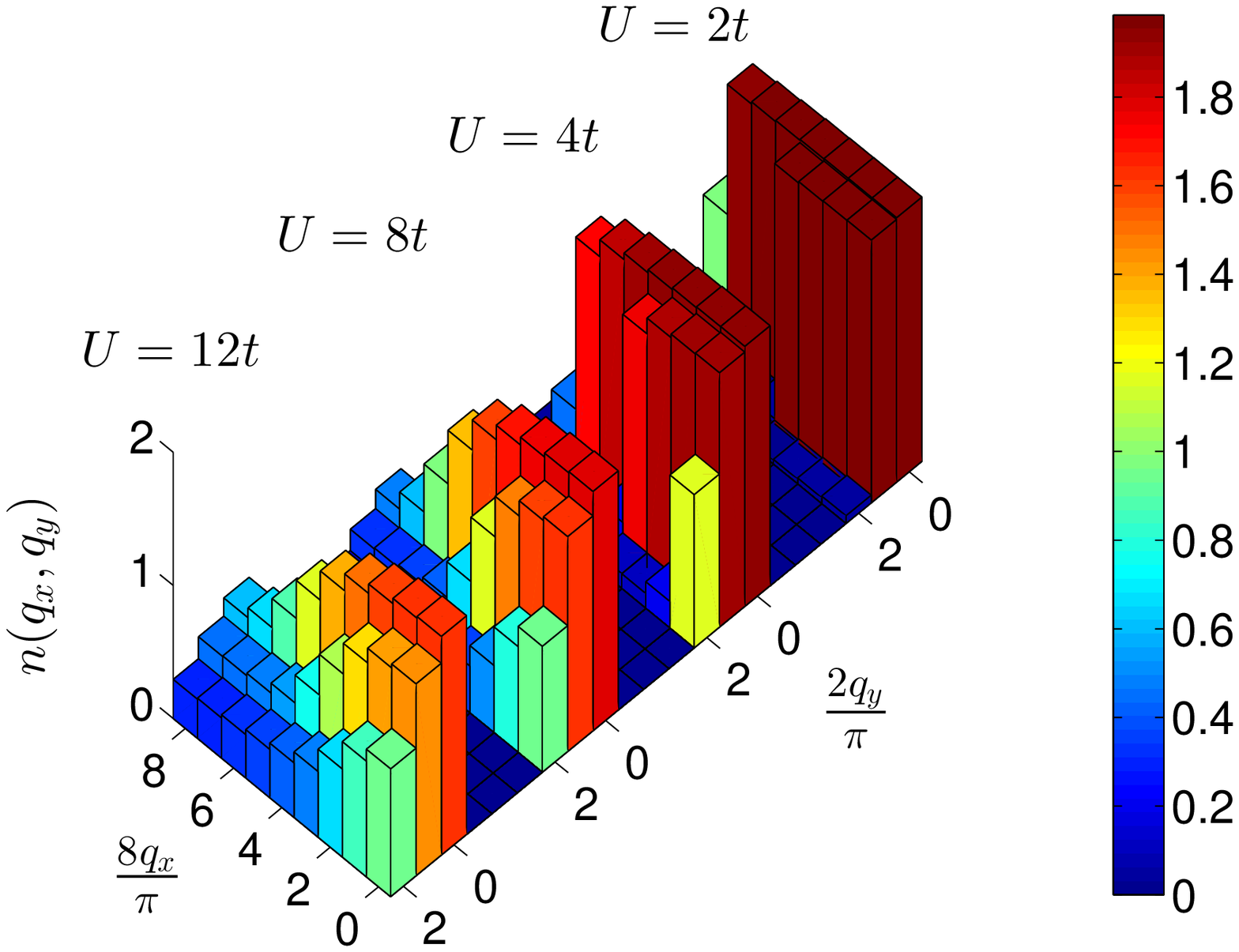} \\
\includegraphics[width=0.525\textwidth]{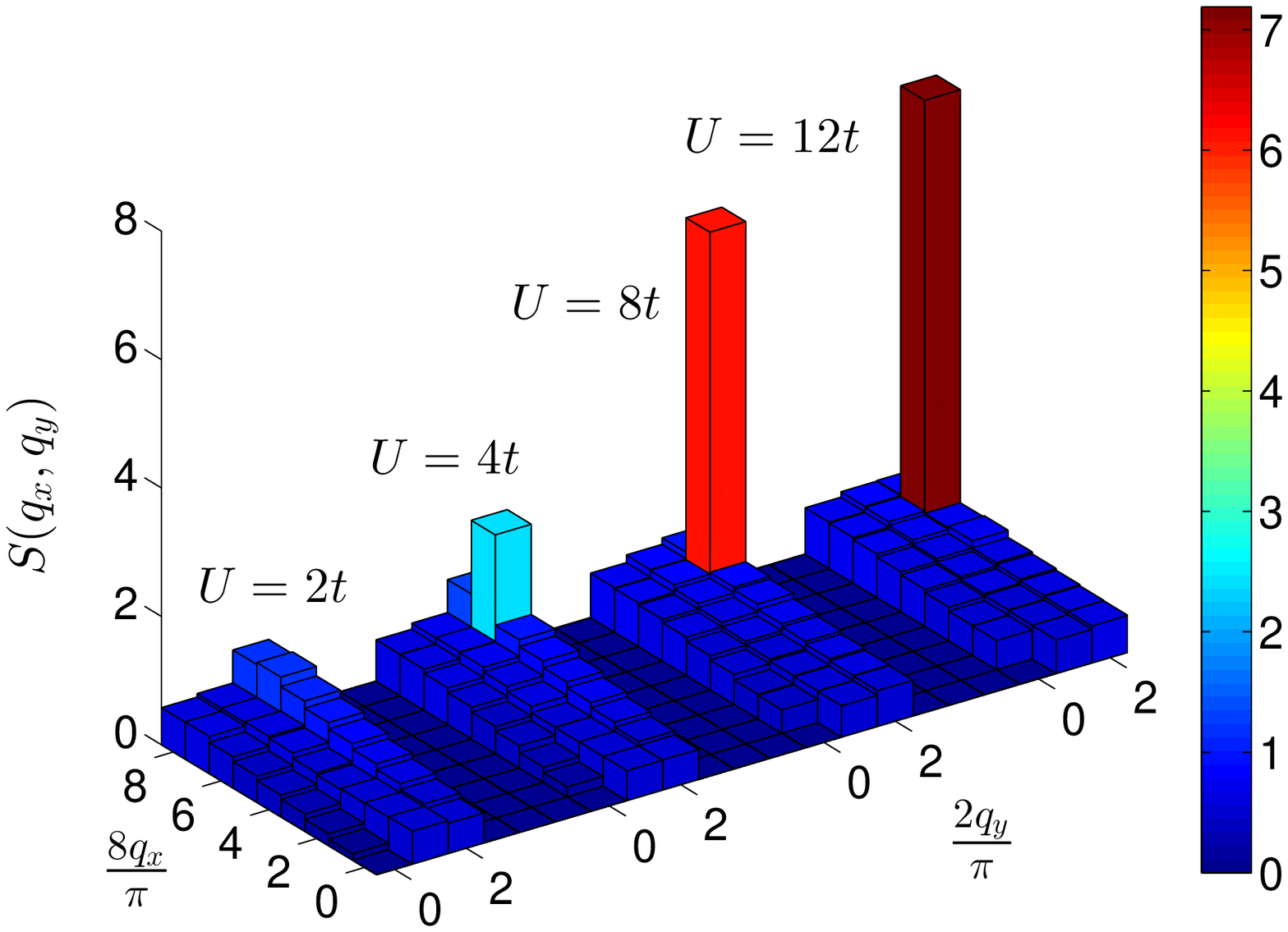} \\
\includegraphics[width=0.525\textwidth]{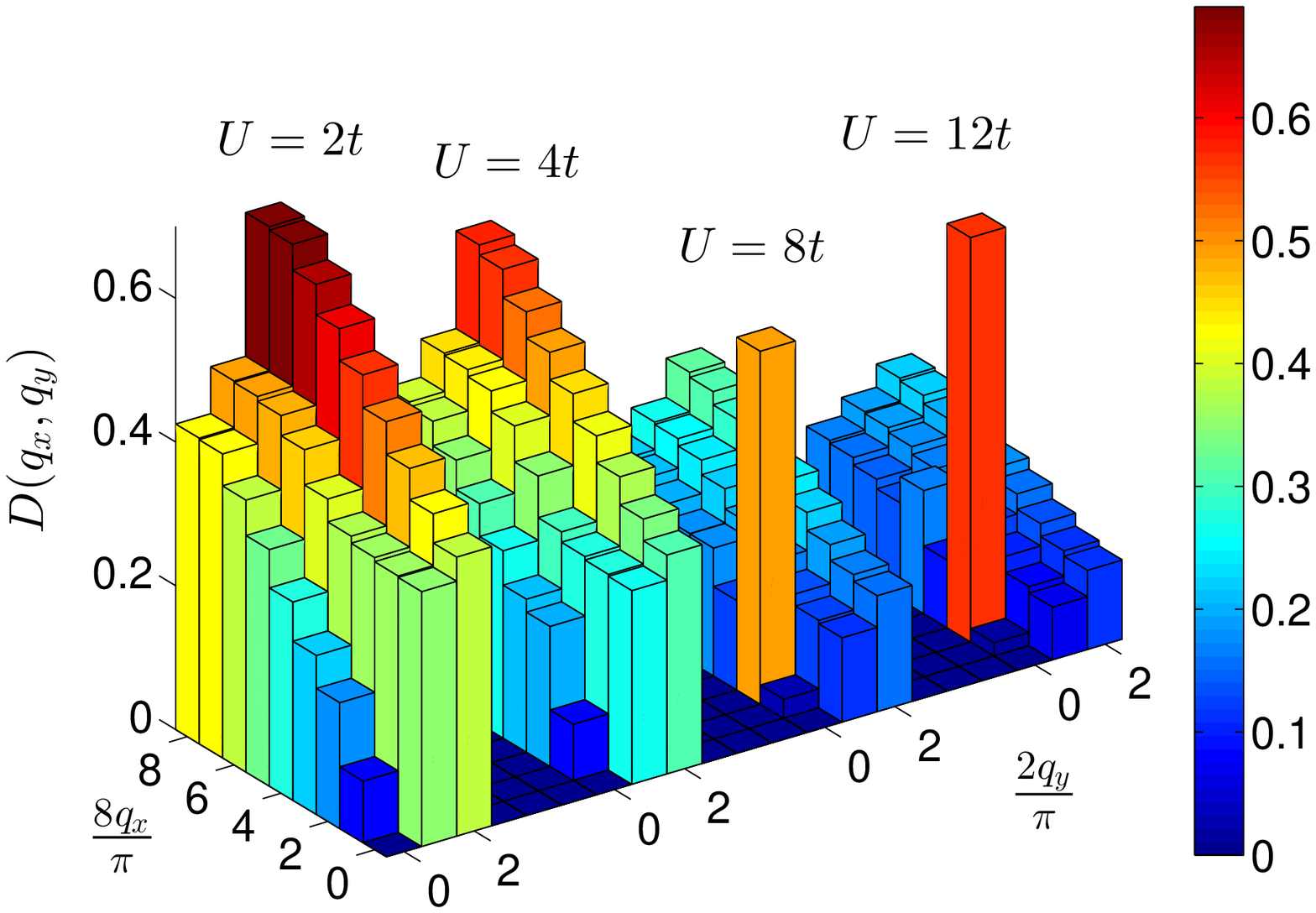}
\caption{(Color online)
Momentum distribution [Eq.(\ref{nkdist})] provided
by UHF-FED calculations, based on n=40 transformations, for the 16 $\times$ 4 lattice with N$_{e}$=56
electrons is shown in the top  panel while the middle and bottom  panels depict the
Fourier transforms of the SSCFs [Eq.(\ref{ss-CF})]  and CCCFs [Eq.(\ref{charge-charge-CF})] in real space.  Only the
upper right quarter of the Brillouin zone is shown.
Results are presented for U=2t, 4t, 8t, and 12t. For more details, see the main text.
}
\label{Fig8}
\end{figure}
%
%

For the same system, we have also computed the SSCFs $S_{m}^{\Theta}({\bf{i}})$ in real space

\begin{eqnarray} \label{ss-CF}
S_{m}^{\Theta}({\bf{i}}) = \frac{4}{3}
\frac{
\langle \phi_{K}^{\Theta} | \hat{\bf{S}}({\bf{i}}) \cdot \hat{\bf{S}}({\bf{1}}) | \phi_{K}^{\Theta} \rangle
}
{
\langle \phi_{K}^{\Theta}  | \phi_{K}^{\Theta} \rangle
}
\end{eqnarray}
where the subindex $m$ accounts for the irreducible representation of the space group
used in the symmetry-projected calculations.
\cite{rayner-Hubbard-1D-FED2013,rayner-Hubbard-1D-FED2014} The
Fourier transforms of the GHF-FED  SSCFs
(black diamonds)
are compared in Fig. \ref{Fig7} with CPQMC results
based on  CASSCF multideterminantal
trial wave functions with symmetries  (orange diamonds)
and ED values (red diamonds).
 \cite{Shiwei-QMC-Symmetry} Results are shown for
the onsite interactions 4t (a), 8t (b) and 12t (c). Regardless of the considered
U values, we observe that the use of states with well defined
symmetry quantum numbers, both within the GHF-FED and CPQMC  schemes,
lead to SSCFs
that agree well with the ED ones.

Calculations have also been performed for the half-filled 6 $\times$ 6 and
8 $\times$ 8 lattices. With only one symmetry-projected GHF configuration
we have obtained the values $S(\pi,\pi)$ = 6.0283 and 9.6164, respectively,
for the Fourier transform of the SSCF at the wave vector ${\bf{q}}=(\pi,\pi)$.
On the other hand, the
MR GHF-FED wave functions already discussed in   Sec. \ref{results-gs-corr-energies}
(i.e., n=120 and  108) lead us to $S(\pi,\pi)$ = 5.8245 and 8.3173, which should be
compared with the
auxiliary-field
 QMC estimates of 5.82(3) and 8.2(2). \cite{aux-QMC-Imada} Therefore, through the
VAP constructive increase of its basis states, Eq.(\ref{FED-state-general}), the FED scheme
improves not only
the ground state energies of the considered lattices, as already discussed in
Sec. \ref{results-gs-corr-energies}, but also captures the most relevant
spin-spin correlations.

The  momentum distributions and
the Fourier transforms of the
ground state
SSCFs and CCCFs
obtained with UHF-FED  calculations (n=40)
for a 16 $\times$ 4 lattice with
N$_{e}$=56 electrons are shown in Fig. \ref{Fig8} for U=2t, 4t, 8t, and 12t. We have tested that the number of basis states
 is large  enough to
already
capture the main features
of the considered quantities and that the corresponding profiles, especially those
for large U values, are not significantly modified
 by further increasing n. At U=2t, the  momentum distribution
 [Eq.(\ref{nkdist})] (top panel)
resembles, to a large extent, the one corresponding to
 a noninteracting Fermi gas
 where the states below the Fermi surface are occupied. With increasing
 onsite repulsions the momentum distributions are smeared out.

The Fourier transforms of the
SSCFs [Eq.(\ref{ss-CF})]  (middle panel) exhibit a broad background for all the considered U values.
For both U=2t and 4t, there exists a weak antiferromagnetic peak at wave vector
${\bf{q}}=(\pi,\pi)$ which already disappears at U=8t. On the other hand, at U=4t, a  peak
can already be seen at ${\bf{q}}=(7\pi/8,\pi)$  which becomes more prominent as the
onsite interaction is increased, signaling the emergence
of inconmesurate spin-spin correlations. Furthermore, we have studied
the CCCFs  given by

\begin{eqnarray} \label{charge-charge-CF}
D_{m}^{\Theta}({\bf{i}}) =
\frac{
\langle \phi_{ K}^{\Theta} | \hat{n}_{\bf{i}} \hat{n}_{\bf{1}} | \phi_{K}^{\Theta} \rangle
}
{
\langle \phi_{K}^{\Theta}  | \phi_{K}^{\Theta} \rangle
}
\end{eqnarray}
where  $\hat{n}_{\bf{i}}= \sum_{\sigma} \hat{n}_{{\bf{i}} \sigma}$. The corresponding
Fourier transforms are shown in the bottom  panel of
Fig. \ref{Fig8}.  In this case, the  quantity
N$_{e}^{2}$/N$_{sites}^{2}$
has been subtracted to take out the trivial peak
at  the origin ${\bf{q}}=(0,0)$. For both U=2t and 4t, they
are broad with little features. However, already at U=8t
a peak appears at wave vector ${\bf{q}}=(\pi/4,0)$
signaling  the development of charge order.
The previous results for SSCFs and CCCFs, are consistent
with the
crossover to a
stripe
regime already anticipated in Sec. \ref{results-gs-corr-energies} in terms of the intrinsic
determinants resulting from the UHF-FED VAP procedure. They agree well
with the ones obtained within the CPQMC approach. \cite{Stripes-Shiwei}

%
%
\begin{figure}
\includegraphics[width=0.505\textwidth]{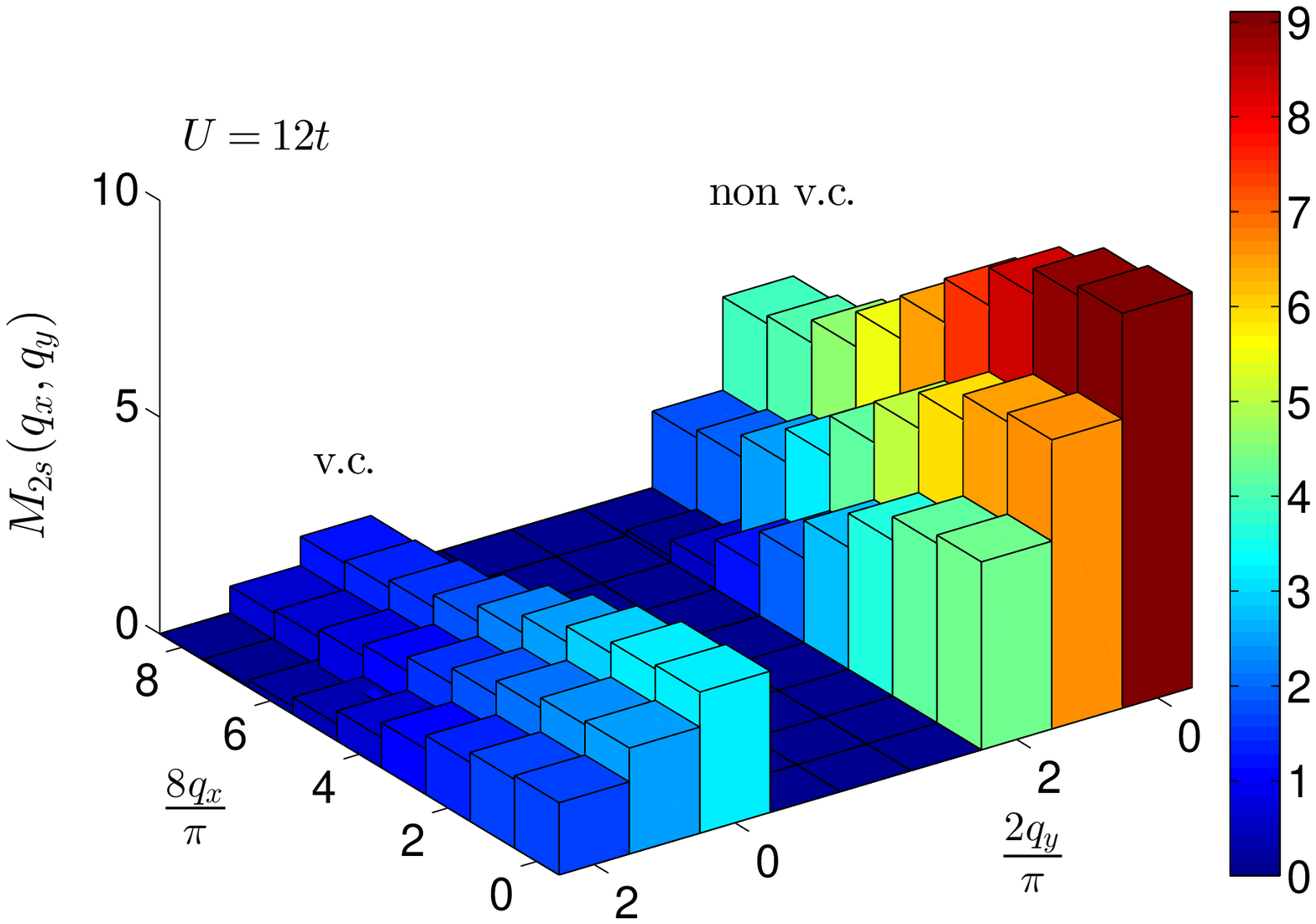} \\
\includegraphics[width=0.505\textwidth]{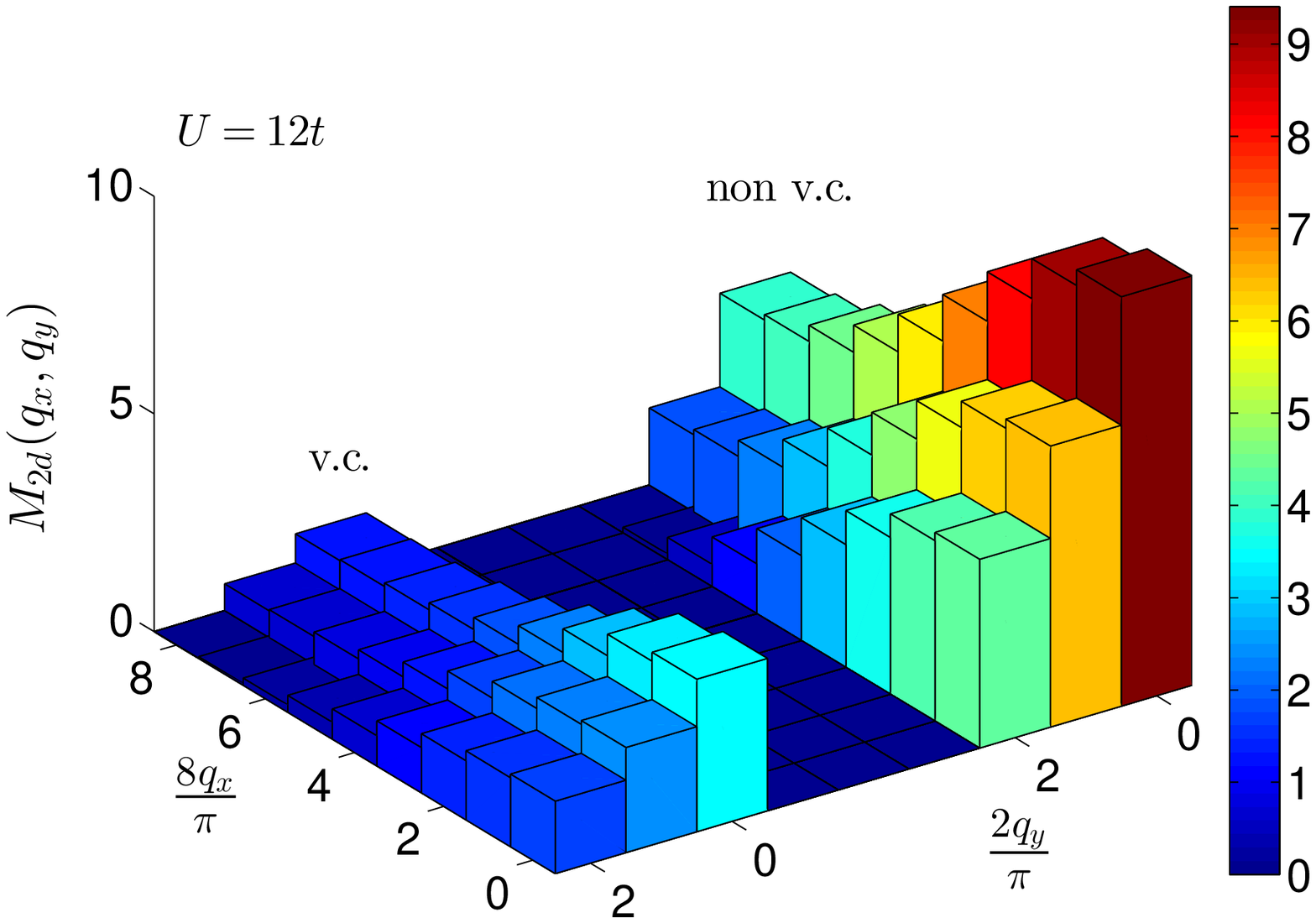}
\caption{(Color online)
Fourier transforms of the
 2s (extended s-wave) and 2d (d$_{x^{2}-y^{2}}$ wave)
PCFs
provided
by UHF-FED calculations, based on n=40 transformations, for the 16 $\times$ 4 lattice with N$_{e}$=56
electrons
are depicted, with (v.c) and without (non v.c) vertex corrections, in the top and bottom
panels. Results are shown for U=12t. For more details, see
the main text.
}
\label{Fig9}
\end{figure}
%
%

Finally, the  PCFs are
 defined in real space as

\begin{eqnarray} \label{pair-pair}
M_{m}^{\Theta}({\bf{i}}) =
\frac{
\langle \phi_{ K}^{\Theta} | \hat{\Delta}_{\bf{1}} \hat{\Delta}_{\bf{i}}^{\dagger} | \phi_{K}^{\Theta} \rangle
}
{
\langle \phi_{K}^{\Theta}  | \phi_{K}^{\Theta} \rangle
}
\end{eqnarray}
with

\begin{eqnarray} \label{pair-pair-delta}
\hat{\Delta}_{\bf{i}}= \sum_{{\bf{R}}} f({\bf{R}})
\Big[
\hat{c}_{{\bf{i}} \uparrow} \hat{c}_{{\bf{i}}+{\bf{R}} \downarrow}
-
\hat{c}_{{\bf{i}} \downarrow} \hat{c}_{{\bf{i}}+{\bf{R}} \uparrow}
\Big]
\end{eqnarray}
where $f({\bf{R}})$ is a form factor
that depends on the pairing mode under consideration. \cite{aux-QMC-Imada}
We have paid attention to  the 2s (extended s-wave) and 2d (d$_{x^{2}-y^{2}}$) pairing modes
defined by the following  form factors

\begin{eqnarray}
f_{2s}({\bf{R}}) =  \delta_{R_{y},0}
\sum_{l=-1,1} \delta_{R_{x},l}
+
\delta_{R_{x},0} \sum_{l=-1,1} \delta_{R_{y},l}
\nonumber\\
f_{2d}({\bf{R}}) =
\delta_{R_{y},0}
\sum_{l=-1,1} \delta_{R_{x},l}
- \delta_{R_{x},0} \sum_{l=-1,1} \delta_{R_{y},l}
\end{eqnarray}

For each mode, we have considered   PCFs  with (v.c) and
without (non v.c) vertex corrections. In the former case, we have
replaced the two-electron density matrix
$\rho({\bf{i}}_{1} \sigma_{1},{\bf{i}}_{2} \sigma_{2},{\bf{i}}_{3} \sigma_{3},{\bf{i}}_{4} \sigma_{4})
=
\langle
\hat{c}_{{\bf{i}}_{1} \sigma_{1}}^{\dagger}
\hat{c}_{{\bf{i}}_{2} \sigma_{2}}^{\dagger}
\hat{c}_{{\bf{i}}_{3} \sigma_{3}}
\hat{c}_{{\bf{i}}_{4} \sigma_{4}}
\rangle
$
by
$\omega({\bf{i}}_{1} \sigma_{1},{\bf{i}}_{2} \sigma_{2},{\bf{i}}_{3} \sigma_{3},{\bf{i}}_{4} \sigma_{4})
= \rho({\bf{i}}_{1} \sigma_{1},{\bf{i}}_{2} \sigma_{2},{\bf{i}}_{3} \sigma_{3},{\bf{i}}_{4} \sigma_{4})
- \rho({\bf{i}}_{4} \sigma_{4},{\bf{i}}_{1} \sigma_{1})
\rho({\bf{i}}_{3} \sigma_{3},{\bf{i}}_{2} \sigma_{2})
+
\rho({\bf{i}}_{3} \sigma_{3},{\bf{i}}_{1} \sigma_{1})
\rho({\bf{i}}_{4} \sigma_{4},{\bf{i}}_{2} \sigma_{2})
$. The quantities
$
\rho({\bf{i}}_{2} \sigma_{2},{\bf{i}}_{1} \sigma_{1}) =
\langle
\hat{c}_{{\bf{i}}_{1} \sigma_{1}}^{\dagger}
\hat{c}_{{\bf{i}}_{2} \sigma_{2}}
\rangle
$
are the one-electron density matrices \cite{rs} and the
mean values $\langle  \dots  \rangle$ are always taken with the FED state Eq.(\ref{FED-state-general}).
With these definitions, a positive vertex-corrected PCF would imply that the effective
electron-electron interaction enhances the considered pairing correlations
with respect to a dressed single-particle picture.
We have plotted the Fourier transforms of the 2s and 2d PCFs,  with (v.c) and
without (non v.c) vertex corrections,  in the top and bottom panels of Fig. \ref{Fig9}, respectively. Results
are shown for U=12t, i.e., in the stripe regime.  As expected, the vertex corrections do not change the profiles of the Fourier
transforms which reflect the pronounced locality of the 2s and 2d pairing correlations
in real space. In good agreement with previous studies, we observe weakly enhanced 2s and 2d
pairing correlations. \cite{Moreo-Scalapino}

\section{Conclusions}
 \label{CONCLU}

In this work, we have applied the FED approach, previously considered only for 1D systems, to the repulsive
2D Hubbard model. Our main goal has been to test the method for
both half-filled and doped lattices. We have compared our results for ground state
and correlation energies with those obtained using other theoretical approximations.
From the results reported in this work and those obtained in our previous studies, \cite{rayner-Hubbard-1D-FED2013,rayner-Hubbard-1D-FED2014}
together with its parallelization properties,  we
 conclude that
 regardless of the dimensionality and/or doping fraction of the considered lattices
 and
 through
its constructive VAP building of a symmetry-projected basis, the FED scheme provides compact
MR correlated wave functions, with well defined quantum numbers, whose quality can be systematically
improved by increasing the number of basis states used in the expansion. In fact, the method could be
seen as a symmetry-projected and variationally-truncated configuration-interaction (CI) approach.
\cite{Carlo-review}
 The key
point is that the hierarchy of the truncation is transferred to a correlated basis of symmetry-projected
multideterminantal (nonorthogonal)
configurations. In this model, it is the Hamiltonian who determines
via the Ritz variational principle [i.e., the Thouless theorem
\cite{rs}
plus the resonon-like Eq.(\ref{HW-1})],
 the relative weight
of each of these nonorthogonal basis states for capturing the most relevant correlations in a given system via
chains  of calculations.

For different lattices sizes, doping fractions, and onsite interactions, we have found
that the intrinsic determinants $| {\cal{D}}^{r} \rangle$
resulting from FED calculations display  a wide variety of structural defects
which encode information about the basic units of quantum
fluctuations. For example, in the case of a 16 $\times$ 4 lattice with a commensurate doping fraction
x=7/8, the varying structure of those defects and the associated charge densities, revealed
the transition to a (fluctuating) stripe regime,
which agrees well with previous results
obtained with an auxiliary-field
QMC approximation. \cite{Stripes-Shiwei}
Similar to the 1D case, \cite{rayner-Hubbard-1D-FED2013,rayner-Hubbard-1D-FED2014} the optimization of the intrinsic
determinants in the presence of the projection operators induces such
structural defects. It is precisely the action of the projection operators
(rotations, translations by one lattice site, etc.) on these defects,
as well as their interaction through the resonon-like Eq.(\ref{HW-1}),
what brings about the substantial correlation energy obtained within the FED scheme
compared to the usual mean-field HF calculations.

We have compared  the FED momentum distributions
and SSCFs with those obtained via the CPQMC approach based
on trial wave functions with well defined symmetries \cite{PaperwithShiwei,Shiwei-QMC-Symmetry}
for the case of a small 4 $\times$ 4 lattice
with N$_{e}$=14 electrons. We conclude that the use of pure spin states leads to a good
agreement between ours, the CPQMC, and ED values. We have then turned our attention to the computation of
SSCFs and CCCFs  for a 16 $\times$ 4 lattice with N$_{e}$=56 electrons. The
corresponding results signal the emergence of incommensurate spin-spin correlations
and the development of charge order  for increasing onsite repulsions that is
consistent with the transition to a stripe regime anticipated in terms
of the structure of the intrinsic Slater states resulting from our
variational strategy. Furthermore, in good agreement with previous
studies, \cite{Moreo-Scalapino} the  (vertex-corrected) PCFs, computed
with the MR FED wave functions, display  a weak
enhancement
of the extended s-wave and d$_{x^{2}-y^{2}}$ pairing modes.

The FED methodology 
has already been quite successful in microscopic nuclear structure theory, \cite{Carlo-review}
but it is still in its first steps in both quantum chemistry
\cite{Carlos-Rayner-Gustavo-FED-molecules,Laimis-paper}
and condensed
matter physics. \cite{rayner-Hubbard-1D-FED2013,rayner-Hubbard-1D-FED2014} We believe that
it is a good candidate for further multidisciplinary bridges between these
research fields.
In the realm of condensed matter physics, a long list of tasks awaits  completion.
Among others, a more
detailed study of the structural defects resulting from our variational strategy
is required including geometries other than square and rectangular ones, i.e., the honeycomb,
triangular, and  kagome lattices. Such  studies could be useful to deepen the understanding
of the basic units of quantum fluctuations in these lattices. Given the prominent role
of defects, their careful classification within the FED approach could also be
useful to further improve the quality of the starting intrinsic configurations used in our (highly nonlinear)
optimizations.

Let us also stress that in this study, we have concentrated on the
repulsive sector  of the model. The FED approach
can also be generalized to include, in addition to
spin and space group, the restoration of the U(1) particle number symmetry on top
of symmetry-broken Hartree-Fock-Bogoliobov states. \cite{Carlo-review} This would allow us to also tackle the
attractive  sector of the Hubbard model.

$
$
$
$

\begin{acknowledgments}

This work was supported
by the Department of Energy, Office of Basic Energy Sciences, Grant No. DE-FG02-
09ER16053. G.E.S. is a Welch Foundation Chair (C-0036).
Some of the calculations in this work
have been performed at the Titan computational facility, Oak
Ridge National Laboratory, National Center for Computational
Sciences, under project CHM048. The authors also acknowledge
a computational grant received from
the National Energy Research Scientific
Computing Center (NERSC) under the project Projected Quasiparticle Theory.

\end{acknowledgments}


\begin{thebibliography}{00}
\bibitem{ACRONY} In our previous work
\cite{PaperwithShiwei}  we have used
 the acronyms SG,S-UHF and SG,S-GHF, respectively.
\end{thebibliography}
\end{document}